\documentclass[a4paper,11pt]{article}
\usepackage{jheppub} 
\pdfoutput=1
\usepackage{amssymb} \usepackage{amsmath} 
\usepackage{mathtools}
\usepackage{graphicx}
\usepackage{epsfig,latexsym,color,xcolor}
\usepackage[normalem]{ulem}
\usepackage{amstext}    
\usepackage{array} 
\usepackage{slashed}
\begin{document}

\def\Giulia{\bf\color{red}}
\def\bef{\begin{figure}}
\def\eef{\end{figure}}
\newcommand{\ans}{ansatz }
\newcommand{\be}[1]{\begin{equation}\label{#1}}
\newcommand{\beq}{\begin{equation}}
\newcommand{\ee}{\end{equation}}
\newcommand{\beqn}[1]{\begin{eqnarray}\label{#1}}
\newcommand{\eeqn}{\end{eqnarray}}
\newcommand{\bd}{\begin{displaymath}}
\newcommand{\ed}{\end{displaymath}}
\newcommand{\mat}[4]{\left(\begin{array}{cc}{#1}&{#2}\\{#3}&{#4}
\end{array}\right)}
\newcommand{\matr}[9]{\left(\begin{array}{ccc}{#1}&{#2}&{#3}\\
{#4}&{#5}&{#6}\\{#7}&{#8}&{#9}\end{array}\right)}

\newcommand{\matrr}[6]{\left(\begin{array}{cc}{#1}&{#2}\\
{#3}&{#4}\\{#5}&{#6}\end{array}\right)}
\newcommand{\cvb}[3]{#1^{#2}_{#3}}
\def\lsim{\raise0.3ex\hbox{$\;<$\kern-0.75em\raise-1.1ex
e\hbox{$\sim\;$}}}
\def\gsim{\raise0.3ex\hbox{$\;>$\kern-0.75em\raise-1.1ex
\hbox{$\sim\;$}}}
\def\abs#1{\left| #1\right|}
\def\simlt{\mathrel{\lower2.5pt\vbox{\lineskip=0pt\baselineskip=0pt
           \hbox{$<$}\hbox{$\sim$}}}}
\def\simgt{\mathrel{\lower2.5pt\vbox{\lineskip=0pt\baselineskip=0pt
           \hbox{$>$}\hbox{$\sim$}}}}
\def\unity{{\hbox{1\kern-.8mm l}}}
\newcommand{\eps}{\varepsilon}
\def\ep{\epsilon}
\def\ga{\gamma}
\def\Ga{\Gamma}
\def\om{\omega}
\def\omp{{\omega^\prime}}
\def\Om{\Omega}
\def\la{\lambda}
\def\La{\Lambda}
\def\al{\alpha}
\def\beq{\begin{equation}}
\def\eeq{\end{equation}}
\newcommand{\MyRed}{\color [rgb]{0.8,0,0}}
\newcommand{\MyGreen}{\color [rgb]{0,0.7,0}}
\newcommand{\MyBlue}{\color [rgb]{0,0,0.8}}
\newcommand{\MyPurple}{\color [rgb]{0.6,0.0,0.6}}
\def\GV#1{{\MyRed [GV: #1]}}
\def\GR#1{{\MyGreen [GR: #1]}}   
\def\DF#1{{\MyBlue [DF: #1]}}
\def\DK#1{{\MyPurple [DK: #1]}}
\newcommand{\sect}[1]{\setcounter{equation}{0}\section{#1}}
\renewcommand{\theequation}{\thesection.\arabic{equation}}
\newcommand{\ov}{\overline}
\renewcommand{\to}{\rightarrow}
\renewcommand{\vec}[1]{\mathbf{#1}}
\def\tm{{\widetilde{m}}}
\def\mcirc{{\stackrel{o}{m}}}
\newcommand{\Dm}{\Delta m}
\newcommand{\dm}{\varepsilon}
\newcommand{\tanb}{\tan\beta}
\newcommand{\nbar}{\tilde{n}}
\newcommand\PM[1]{\begin{pmatrix}#1\end{pmatrix}}
\newcommand{\up}{\uparrow}
\newcommand{\down}{\downarrow}
\newcommand{\refs}[2]{eqs.~(\ref{#1})-(\ref{#2})}
\def\omE{\omega_{\rm Ter}}
\newcommand{\eqn}[1]{eq.~(\ref{#1})}
%

\newcommand{\DSUSY}{{SUSY \hspace{-9.4pt} \slash}\;}
\newcommand{\DCP}{{CP \hspace{-7.4pt} \slash}\;}
\newcommand{\mc}{\mathcal}
\newcommand{\gr}{\mathbf}
\renewcommand{\to}{\rightarrow}
\newcommand{\gtc}{\mathfrak}
\newcommand{\wh}{\widehat}
\newcommand{\br}{\langle}
\newcommand{\kt}{\rangle}
\newcommand{\ra}{\rightarrow}


\def\lsim{\mathrel{\mathop  {\hbox{\lower0.5ex\hbox{$\sim$}
\kern-0.8em\lower-0.7ex\hbox{$<$}}}}}
\def\gsim{\mathrel{\mathop  {\hbox{\lower0.5ex\hbox{$\sim$}
\kern-0.8em\lower-0.7ex\hbox{$>$}}}}}

\def\nn{\\  \nonumber}
\def\de{\partial}
\def\brf{{\mathbf f}}
\def\bbf{\bar{\bf f}}
\def\bF{{\bf F}}
\def\bbF{\bar{\bf F}}
\def\bA{{\mathbf A}}
\def\bB{{\mathbf B}}
\def\bG{{\mathbf G}}
\def\bI{{\mathbf I}}
\def\bM{{\mathbf M}}
\def\bY{{\mathbf Y}}
\def\bX{{\mathbf X}}
\def\bS{{\mathbf S}}
\def\bb{{\mathbf b}}
\def\bh{{\mathbf h}}
\def\bg{{\mathbf g}}
\def\bla{{\mathbf \la}}
\def\bmu{\mathbf m }
\def\by{{\mathbf y}}
\def\bmu{\mbox{\boldmath $\mu$} }
\def\bsig{\mbox{\boldmath $\sigma$} }
\def\bunity{{\mathbf 1}}
\def\cA{{\cal A}}
\def\cB{{\cal B}}
\def\cC{{\cal C}}
\def\cD{{\cal D}}
\def\cF{{\cal F}}
\def\cG{{\cal G}}
\def\cH{{\cal H}}
\def\cI{{\cal I}}
\def\cL{{\cal L}}
\def\cN{{\cal N}}
\def\cM{{\cal M}}
\def\cO{{\cal O}}
\def\cP{{\cal P}}
\def\cR{{\cal R}}
\def\cS{{\cal S}}
\def\cT{{\cal T}}
\def\eV{{\rm eV}}

\newcommand{\JJ}{J\bar{J}}
\newcommand{\Pexp}{\mathcal{P}\exp}
\newcommand{\tr}{\mathrm{tr}}

%

\hfill \hbox{CERN-TH-2024-055}
\vskip 2.0cm

\title{Baryon-number - flavor  separation \\
 in the topological expansion of QCD}

\author[a]{David Frenklakh}
\author[a,b]{, Dmitri Kharzeev}
\author[c,d]{, Giancarlo Rossi}
\author[e,f]{and Gabriele Veneziano}
\affiliation[a]{Center for Nuclear Theory, Department of Physics and Astronomy, Stony Brook University,\\Stony Brook, NY 11974-3800, USA}
\affiliation[b]{Department of Physics, Brookhaven National Laboratory, Upton, NY 11973-5000, USA}
\affiliation[c]{Dipartimento di Fisica, Universit\`a di  Roma
  ``Tor Vergata''  INFN, Sezione di Roma \\ ``Tor Vergata'', via della Ricerca Scientifica - 00133 Roma, Italy}
\affiliation[d]{Centro Fermi - Museo Storico della Fisica e Centro Studi e Ricerche E.\ Fermi, \\ via Panisperna 89a, 00184 Roma, Italy}
\affiliation[e]{Theory Department, 
CERN, CH-1211,  Geneva 23, Switzerland}
\affiliation[f]{Coll\'ege de France, 11 place M. Berthelot, 75005 Paris, France}

\emailAdd{david.frenklakh@stonybrook.edu, dmitri.kharzeev@stonybrook.edu, rossig@roma2.infn.it, gabriele.veneziano@cern.ch}

\abstract{Gauge invariance of QCD dictates the presence of string junctions in the wave functions of baryons \cite{Rossi:1977cy}. In high-energy inclusive processes, these baryon junctions have been predicted to induce the separation of the flows of baryon number and flavor \cite{Kharzeev:1996sq}. In this paper we 
describe this phenomenon using 
the analog-gas model of multiparticle production proposed long time ago by Feynman and Wilson \cite{Wilson:1970zzb} and adapted here to accommodate the topological expansion in QCD \cite{Veneziano:1974fa,Veneziano:1976wm}. In this framework, duality arguments suggest   the existence of two degenerate junction-antijunction glueball Regge trajectories of opposite $\cal{C}$-parity with intercept close to 1/2. The corresponding results for the energy and rapidity dependence of baryon stopping are in reasonably good agreement with recent experimental findings from STAR and ALICE experiments. We show that accounting for correlations between the fragmenting strings further improves agreement with the data, and outline additional experimental tests of our picture at the existing (RHIC, LHC, JLab) and future (EIC) facilities.}


\maketitle

\newpage
\section{Introduction}

In QCD the wave functions of asymptotic states (hadrons) have to be gauge-invariant. For a meson composed by a quark located at point $x_2$ and an antiquark at point $x_1$, the gauge invariance can be enforced by introducing the path-ordered exponential operator
\begin{eqnarray}\label{WL}
U({\cal C}(x_1,x_2)) \equiv {P} \exp\left({i g \int_{\mathcal{C}(x_1,x_2)} A_\mu(x) dx^\mu}\right)  \end{eqnarray}
constructed from the gluon field $A_\mu \equiv t^a A_\mu^a$, yielding the meson wave function
\begin{eqnarray}\label{meson}
    M({\cal C}(x_1,x_2)) =  \bar q(x_1) U(\mathcal{C}(x_1,x_2)) q(x_2).
    \end{eqnarray}
    Here $\cal C$ denotes a path going from $x_1$ to $x_2$. For baryons, the unique way to construct a gauge-invariant wave function is to join three string operators at a junction, as proposed in \cite{Rossi:1977cy}: 
\begin{eqnarray}\label{baryon}
    \hspace{-.4cm}B({\cal C}_1,{\cal C}_2,{\cal C}_3) = \epsilon^{ijk} \left[U({\cal C}_1(x,x_1)) q(x_1)\right]_i \left[U({\cal C}_2(x,x_2)) q(x_2)\right]_j
    \left[U({\cal C}_3(x,x_3)) q(x_3)\right]_k .
    \end{eqnarray}
The string operator $U({\cal C}(x,x_n))$ acting on a quark field located at space-time point $x_n$ makes it transform under gauge transformations as a field at point $x$. The antisymmetric tensor then makes a color-singlet and gauge-invariant state out of the three quarks plus the junction. 

 At strong coupling, the string operators describe strings connecting the quarks to a common ``baryon junction", which thus becomes, along with the three quarks, a fourth constituent of the baryon \cite{Rossi:1977cy} \footnote{We should stress that in such a constituent-parton picture the junction itself is in a non-trivial color representation such as to compensate for the one of the three valence quark system.},\footnote{For heavy/static quarks the junction naturally emerges also in the context of the AdS/CFT correspondence \cite{Witten:1998xy} and in lattice QCD at strong coupling \cite{Rossi:2016szw}. Its appearence at the so-called Fermat-Torricelli point has been nicely confirmed by lattice-QCD calculations  \cite{Bissey:2006bz}.}.  Baryonic (anti)junctions ($\bar{J}$)$J$ give rise to new exotic hadrons, including baryonium-like $J-\bar{J}$ states with additional two, one, or no $q-\bar{q}$ constituent quark-antiquark pairs, 
 penta-quarks, dibaryons, and other exotics \cite{Rossi:1977cy,Montanet:1980te}.
 It was stressed in \cite{Rossi:1977cy} (see also \cite{Rossi:2016szw}) that duality diagrams involving baryons and anti-baryons are ambiguous unless the flow of the junction, indicating the flow of baryon number, is shown together with the flow of flavor associated with the valence quarks. Regge trajectories containing $J-\bar{J}$ states provide new contributions to high-energy amplitudes of exclusive processes involving baryons and antibaryons \cite{Rossi:1977cy,Montanet:1980te}. 

\vskip0.3cm

In high energy inclusive processes involving baryons, the existence of junctions was predicted to lead to the separation of flows of baryon number and flavor \cite{Kharzeev:1996sq}~\footnote{This phenomenon is somewhat analogous to the spin-charge separation in condensed matter physics (see e.g. \cite{Girvin_Yang_2019}), which also has a topological origin.}. The origin of this phenomenon is easy to understand. At strong coupling, the string operators in (\ref{baryon}) describe strings inside the baryon. In high energy baryon collisions, the valence quarks that carry large Bjorken momentum fraction $x$ end up in the fragmentation region of the process, but the gluonic junction can be more easily stopped in the central rapidity region. The strings that connect the junction to the valence quarks then break by producing additional quark-antiquark pairs, but the baryon is always reconstructed around the junction. Its valence quark content is however not correlated with the one of the original baryon participating in the collision -- so the flows of baryon number and flavor are uncorrelated and separated in rapidity. The  computation of inclusive baryon stopping cross section was performed in \cite{Kharzeev:1996sq} within the Mueller-Kancheli generalized optical theorem approach \cite{Mueller:1970fa,Kancheli:1970gt}, using for the intercept of the leading junction-antijunction quarkless $J_0$ trajectory the value $\alpha_{\mathbb{J}_0} = 0.5$ originally proposed in  \cite{Rossi:1977cy}. The produced baryons are distributed in the c.m.s. rapidity $y_f$ according to \cite{Kharzeev:1996sq}:
\beq
\frac{d N}{d y_f} \propto e^{(\alpha_{\mathbb{J}_0} + \alpha_{\mathbb{P}}-2)Y/2}[e^{(\alpha_{\mathbb{P}}-\alpha_{\mathbb{J}_0})y_f} + e^{(\alpha_{\mathbb{J}_0}-\alpha_{\mathbb{P}})y_f} ], \label{eq:rapidity_dist}
\eeq 
where $Y$ is the rapidity separation between the beams.

\begin{figure}
\centering
\includegraphics[width=0.49\linewidth]{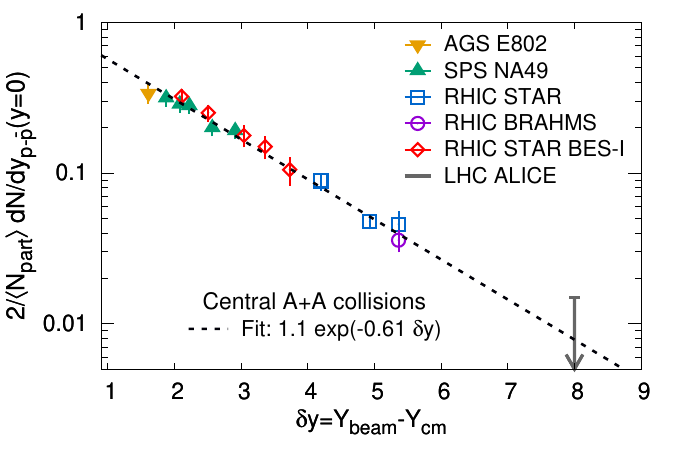}
\includegraphics[width=0.5\linewidth]{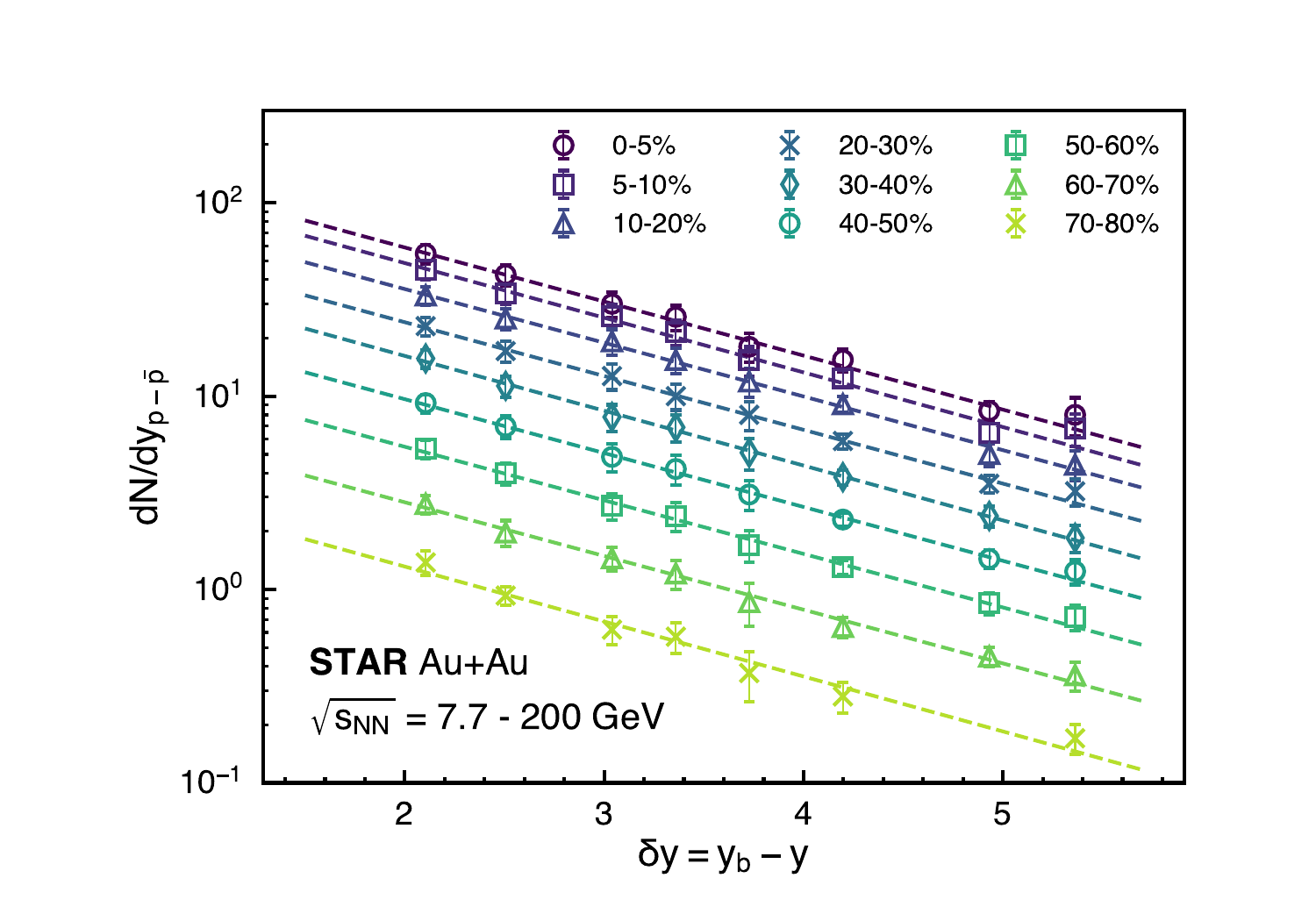}
    \caption{Left panel: compilation of data from various experiments on the rapidity slope parameter that is related to the intercept of $J_0$ trajectory by the means of Eq. (\ref{eq:rapidity_dist}). Right panel: rapidity distribution of baryons from the Beam Energy Scan at RHIC. Evidently, the slope is almost independent of centrality in agreement with baryons stopping mediated by $J_0$ exchange. The centrality-averaged fit to the slope is about $0.65\pm 0.1$ \cite{Lewis:2022arg}. In both panels $\delta y = Y/2$ is the beam rapidity in c.m.s. Both figures are reproduced from \cite{Lewis:2022arg} with kind permission of The European Physical Journal (EPJ).}
    \label{fig:STAR}
\end{figure}

\vskip0.3cm
Recently baryon stopping was studied at RHIC in $AA$ and $\gamma A$ (through ultraperipheral processes) collisions \cite{Lewis:2022arg}. The beam rapidity dependence of the number of stopped baryons is shown in Fig.\ (\ref{fig:STAR}), reproduced from \cite{Lewis:2022arg}. In Eq. (\ref{eq:rapidity_dist}) the slope of the beam rapidity dependence is given by $(\alpha_{\mathbb{J}_0} + \alpha_{\mathbb{P}}-2) = (\alpha_{\mathbb{J}_0} + \Delta - 1)$  where we used $\Delta = \alpha_{\mathbb{P}} - 1 \simeq 0.08$, as fixed by the ``soft" Pomeron intercept $\alpha_{\mathbb{P}} 
\simeq 1.08$ \cite{ParticleDataGroup:2022pth}. If one uses the value of $\alpha_{\mathbb{J}_0} \simeq 0.5$, as originally suggested in \cite{Rossi:1977cy}, the beam rapidity dependence becomes $e^{-0.42\ Y/2}$, to be contrasted with $e^{-0.65\ Y/2}$ from the Beam Energy Scan at RHIC, see the right panel of Fig. (\ref{fig:STAR}). Thus, the agreement between the data and the prediction made with the original value of $\alpha_{\mathbb{J}_0}$ is reasonable. However, with the increased precision of the data it is necessary to revisit the prediction for the $\alpha_{\mathbb{J}_0}$ intercept. This is what we attempt to do in this paper.

\vskip0.3cm

Because the processes of baryon stopping are dominated by small or moderate momentum transfer, they have to be described in the strong coupling domain where the topological expansion of QCD \cite{Veneziano:1973nw,Veneziano:1974fa} can be used as a guiding principle. It is important to develop an approach to high-energy baryon interaction that combines known general features of inclusive processes with the properties of the topological expansion of QCD. For this purpose, it is useful to refer to  an approach to high energy multi-hadron production processes proposed long time ago by Feynman and Wilson.
\vskip0.3cm

In the early seventies, after Feynman's introduction of the concept of inclusive reactions \cite{Feynman:1969ej} (see also \cite{Feynman:1969wa}) and his rediscovery of the Amati-Fubini-Stanghellini scaling in the multiperipheral model \cite{Amati:1962nv} (see also \cite{Wilson:1963tga}), Feynman himself and Ken Wilson (\cite{Wilson:1970zzb}  and private communication cited therein) introduced an interesting analog model of multiparticle interactions assimilating the generating function(al) of inclusive cross sections to the grand-canonical partition function of a gas in a finite volume.

 One obstacle  encountered in the Feynman-Wilson-gas (FWG) formulation was to define and isolate the kind of inclusive cross section (including the simplest of them, the total cross section) to which the FWG formulation could be applied. In particular, Ref. \cite{Wilson:1970zzb}  argues that diffractive processes (diffractive scattering and dissociation) should be taken out before defining the 
 gas analog and its properties.
 
 Large-$N$ expansions, either in the context of the hadronic string or of QCD, can help precisely in formulating the FWG model on a much firmer theoretical basis. However, since multiparticle production is an essential element of the FWG model, the large-$N_{c}$ expansion by 't Hooft \cite{tHooft:1973alw}, in which quark loops are suppressed, cannot be used. Rather, one can appeal to the topological expansion (TE) formulated by one of us, either in the context of the dual resonance model \cite{Veneziano:1974fa} or, even better, of QCD \cite{Veneziano:1976wm}, where the large-$N$ limit is taken at fixed $N_f/N_{c}$.
 Also instrumental for justifying some assumptions is the use of the Mueller-Kancheli approach \cite{Mueller:1970fa,Kancheli:1970gt} to  compute the high-energy behavior of inclusive cross sections through appropriately generalized optical theorems.
 \vskip0.3cm 

The plan of the paper is as follows. 
In Section \ref{sec:PFWG} we concentrate our attention on the planar approximation and its non-linear, exact bootstrap constraints, obtaining a relation between the $q\bar{q}$ leading Reggeon intercept and the ``pressure" of a one-species FWG.
In Section \ref{sec:FWGC} we turn our attention to the cylinder topology which is supposed to describe the bare soft Pomeron appearing at the next to leading order in the TE. Its description is in terms of a two-species FWG. We are thus able to connect the distance of the Pomeron intercept $\alpha_\mathbb{P}$ from 1 to the cross correlations between the two species. 
In Section \ref{sec:FWGB}  we move to the case of $B\Bar{B}$ annihilation into mesons which is suitably described in terms of a three-species FWG. We provide estimates of the intercepts of Regge trajectories corresponding to mesons with a junction-antijunction pair that control different inclusive annihilation cross sections. 
In Section \ref{sec:glueball} we use the estimated $J_0$ Regge trajectory to predict the spectrum of $J_0$ glueballs that can be looked for in lattice QCD calculations and we propose an expression of the glueball operator that can be used for that purpose. 
Section \ref{sec:TFBN} discusses the main observable consequences of our framework: it deals first, separately, with flavor and baryon-number transport in various processes (meson-meson, meson-baryon, baryon-baryon, baryon-antibaryon scattering), it then extends the treatment to the combined flavor--baryon-number rapidity distribution, and finally ends with a discussion of the inclusive  distribution of $B\Bar{B}$ pair production, as a function of their relative rapidity. 
In Section \ref{sec:exp} we propose several independent experimental tests to further examine the role of baryon junctions in high-energy processes. 
In Appendix A we outline (our understanding of) the FWG model in the case of inclusive cross sections as functions of rapidity (i.e. after having integrated over the transverse phase space).


\section{The planar Feynman-Wilson gas (FWG) and the Reggeon intercept}
\label{sec:PFWG}
\setcounter{equation}{0}

In the planar limit for meson-meson scattering the FWG model consists of just one species of mesons  
for which one can define the grand-canonical partition function 
\beqn{plSigma}
\Sigma_{pl}(z) = \frac{1}{\sigma^{pl}_t} \sum_{n\geq 2} z^n \sigma^{pl}_n \quad\Rightarrow \quad\Sigma_{pl}(1) =1
\eeqn
where we have normalized the individual exclusive cross sections $\sigma^{pl}_n$ to the total planar cross section $\sigma^{pl}_t$ and we omitted to indicate that all these quantities depend on the center-of-mass  energy $\sqrt{s}$.
Assuming short-range correlations we may also write (see Appendix A):
\beqn{plSigma1}
&\Sigma_{pl}(z) \equiv \exp\left(Y p(z) \right)= \exp\left(Y\sum_{m \ge1} c_m \frac{(z-1)^m}{m!}\right) ~;~\nonumber \\
& p(1)=0 ~,~ p'(1) Y = c_1 Y = \langle n \rangle~,~ p''(1) Y =  c_2 Y =  \langle n(n-1) \rangle - \langle n \rangle^2~,~ \nonumber \\
 & p'''(1) Y =  c_3 Y =  \langle n(n-1)(n-2) \rangle - \langle n \rangle^3 - 3 c_1 c_2 Y^2 ~,~ \nonumber \\
 & p''''(1) Y =  c_4 Y =  \langle n(n-1)(n-2)(n-3) \rangle - \langle n \rangle^4 - 6 c_1^2 c_2 Y^3 - 4 c_1 c_3 Y^2 - 3c_2^2 Y^2,    \dots
\eeqn
where $p(z)$ plays the role of the pressure \footnote{As explained in Appendix A, $p(z)$ is  insensitive to any $Y$-independent prefactor needed   to  make $\Sigma$ approach a finite limit as $z \to 0$. Also,  we will only be interested in pressure differences which do not depend on the overall normalization of $\Sigma_{pl}$. This is why our pressure goes to zero at $z=1$ while in a real gas it would make more sense to have it vanish at $z=0$.}, as a function of fugacity $z$, and $Y \sim \log (s/\langle p_{\perp}^2\rangle)$ is the rapidity (i.e.\ after having integrated over the transverse momenta, assumed to be strongly cut-off) ``volume'' occupied by the gas.
It is a crucial fact, implied by Regge-pole factorization/dominance, that in (\ref{plSigma}) a single power of $Y$ can be factored out, which corresponds to the assumption of short range correlations.

\begin{table}[!h]
\begin{center}
\begin{tabular}{|c||c||c|}
\hline
$MM$-planar & $MM$-cylinder & $B\bar B$-annihilation  \\
\hline\hline
$\sigma_{excl}^{pl}   \sim s^{2\alpha_\mathbb{R}-2}$  &   & $\sigma_{excl}^{ann}   \sim s^{2\alpha_{\mathbb{B}}-2}$  \\
\hline
&   $\sigma^{cyl}(n,0)   \sim s^{\alpha_\mathbb{R}-1}$  & $\sigma^{ann} (n_1,0,0)  \sim s^{\alpha_{\mathbb{J}_4}-1}$  \\
\hline
&     & $\sigma^{ann} (n_1,n_2,0)  \sim s^{\alpha_{\mathbb{J}_2}-1}$  \\
\hline
$\sigma_t^{pl}   \sim s^{\alpha_\mathbb{R}-1}$  &  $\sigma_t^{cyl}   \sim s^{\alpha_\mathbb{P}-1}$   & $\sigma_t^{ann} =\sigma^{ann} (n_1,n_2,n_3)  \sim s^{\alpha_{\mathbb{J}_0}-1}$  \\
\hline
\end{tabular}
\end{center}
\caption{Leading Regge behavior of cross-sections in different processes.}
\label{tab:tab1}
\end{table}

From standard Regge-pole dominance at high energy (see Table \ref{tab:tab1}), and from the planar optical theorem (see Fig. (\ref{f1})), it follows that, as $z \ra 0$:
\beqn{alphaR}
\Sigma_{pl}(z) \ra \frac{\sigma^{pl}_{excl}}{\sigma^{pl}_t} \sim s^{\alpha_\mathbb{R}-1}  \Rightarrow - p(0) \!=\!  1\!-\! \alpha_\mathbb{R}(0)  \!=\! \sum_m c_m \frac{(-1)^{m+1}}{m!} \!=\! \frac{\langle n \rangle}{Y}\! \!- \frac{c_2}{2} \!+\! \dots
\eeqn
This planar bootstrap result connects the Reggeon intercept to multiparticle correlations. In the case of vanishing correlations (i.e.\ of a Poisson distribution) one recovers a very old result by Chew and Pignotti \cite{Chew:1968fe} based on the simplest multiperipheral model, namely
\beqn{Poisson}
  \alpha_\mathbb{R}(0)  = 1- \frac{\langle n \rangle}{Y} 
\eeqn
Hereafter, to simplify the notation, we will just indicate by $\alpha_i$ the intercept $\alpha_i(0)$ of a generic Regge trajectory.
\begin{figure}[t]
\centering
\includegraphics[scale=0.5]{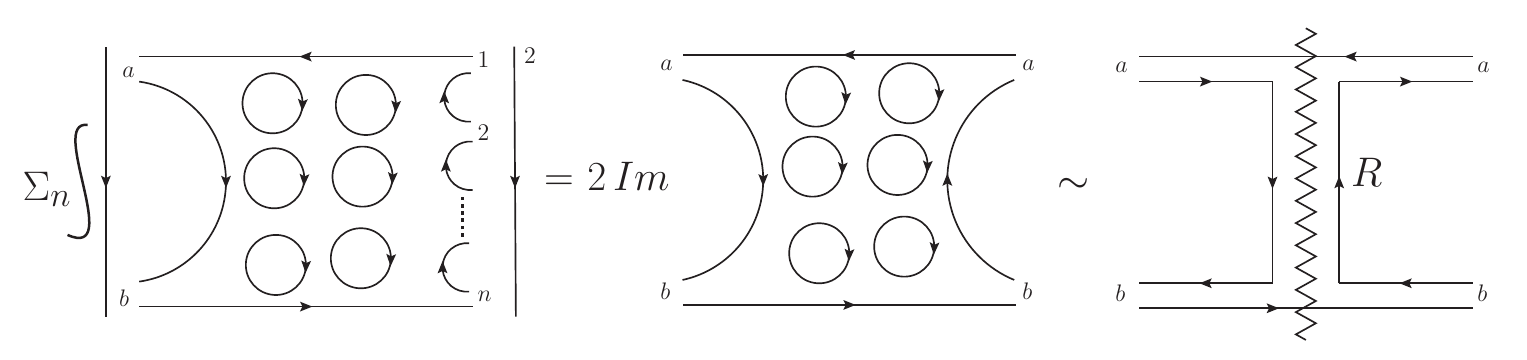}
\caption{From left to right: optical theorem for the total planar cross section $\sum_X \sigma(a+b \ra X)$ and its high-energy limit described by the leading $q\bar{q}$ Regge-pole exchange.}
\label{f1}
\end{figure}

\section{FWG at the cylinder level} 
\label{sec:FWGC}

\setcounter{equation}{0}
At the cylinder level the FWG consists of two species that we shall call ``right"($R$) and ``left" ($L$) according to the picture in Fig. (\ref{f2}).
We  define accordingly:
\beqn{CylSigma}
\Sigma_{cyl}(z_R,z_L) = \frac{1}{\sigma^{cyl}_t} \sum_{n_R+n_L \ge 2} z_R^{n_R} z_L^{n_L} \sigma^{cyl}(n_R,n_L) \Rightarrow \Sigma_{cyl}(1,1) =1
\eeqn
where we chose to normalize the individual exclusive cross-sections $\sigma^{cyl}(n_R,n_L)$ by the total cylinder cross-section $\sigma^{cyl}_t$.
\begin{figure}[t]
\centering
\includegraphics[scale=0.5]{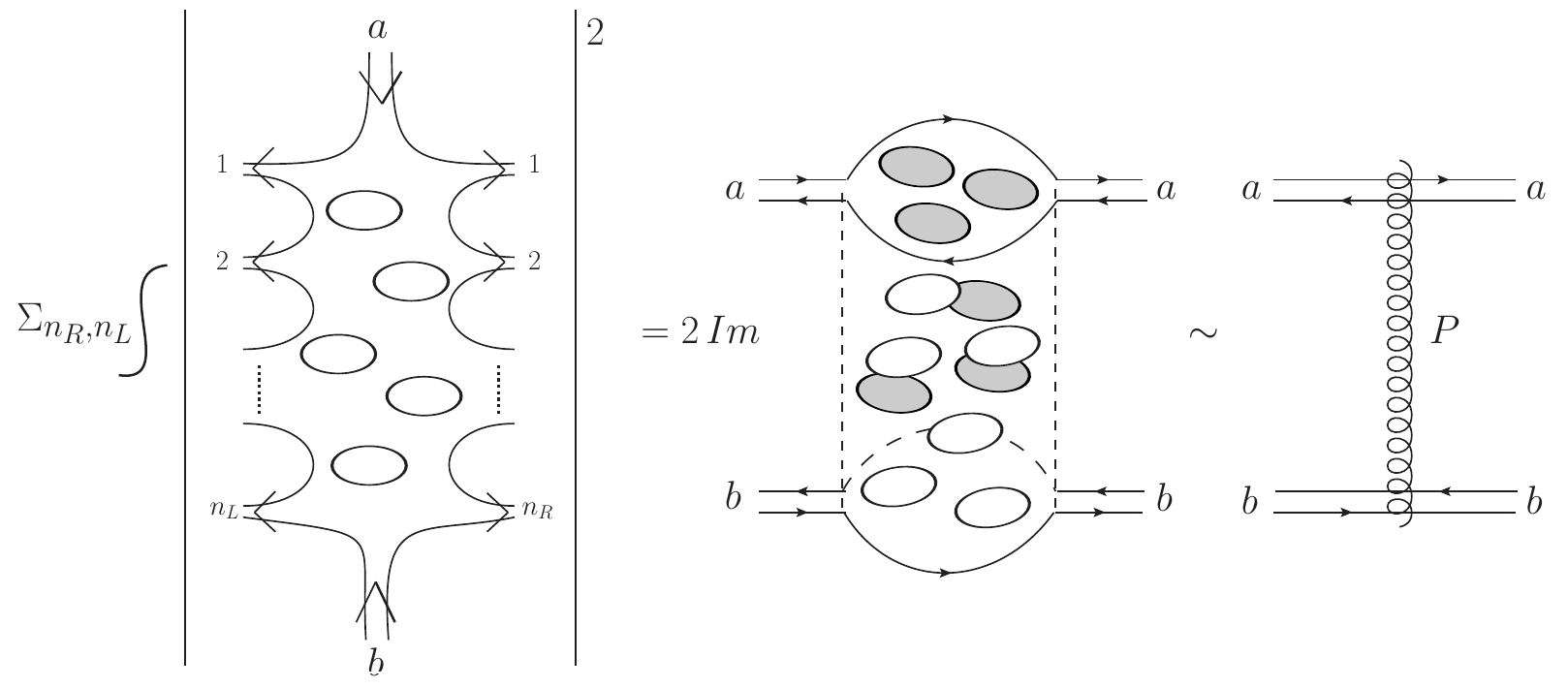}
\caption{Total cylinder cross section seen as a two-species FWG and its high energy limit described by the (bare) Pomeron exchange.}
\label{f2}
\end{figure}
Assuming short-range correlations we now have, in place of (\ref{plSigma1}): 
\beqn{CylSigma1}
&\Sigma_{cyl}(z_R, z_L) \equiv \exp\left(Y p(z_R,z_L) \right) = \exp\left(Y \sum_{m_R+m_L \ge 1} c(m_R,m_L) \frac{(z_R-1)^{m_R} (z_L-1)^{m_L} }{m_R! m_L!}\right)   ~;~ \nonumber \\
&   c(1,0)Y  = \langle n_R \rangle~,~   c(0,1)Y = \langle n_L \rangle~,~  c(1,1)Y =  \langle n_R n_L \rangle - \langle n_R \rangle \langle n_L \rangle~,~ \dots
\eeqn
It is convenient to separate out in (\ref{CylSigma1}) the terms with $n_R=0$ and  $n_L=0$ and write:
\beqn{CylSigma2}
&&p(z_R,z_L) = p_R(z_R) + p_L(z_L) + C_{RL}(z_R,z_L) \nonumber\\
&&p_R(z_R) = \sum_{m_R \ge 1}  c(m_R, 0) \frac{(z_R-1)^{m_R}  }{m_R!}\, ,\qquad p_L(z_L) = \sum_{m_L \ge 1} c(m_L, 0) \frac{(z_L-1)^{m_L}  }{m_L!} \nonumber\\
 &&C_{RL}(z_R,z_L) =  \sum_{m_R, m_L \ge 1}  c(m_R,m_L) \frac{(z_R-1)^{m_R} (z_L-1)^{m_L} }{m_R! m_L!} 
 \eeqn
 One readily finds 
 \beqn{CylSigma3}
&&   p(1,1) = p_R(1) = p_L(1) = C_{RL}(1,z_L) = C_{RL}(z_R,1) =0 ~.
\eeqn

\subsection{The Pomeron Intercept}
\label{sec:PI}

In order to express the (bare)-Pomeron intercept $\alpha_\mathbb{P}$ in terms of the Reggeon intercept $ \alpha_\mathbb{R}$ and the above correlators we note that 
\beqn{alphaP}
p(1,0) + p(0,1) - p(0,0) +  C_{RL}(0,0)=0 \, ,
\eeqn
where we have used the equations (\ref{CylSigma2}) as well as the  relations in (\ref{CylSigma3}). On the other hand, recalling the results summarized in Table \ref{tab:tab1}, we have $p(1,0) = p(0,1) = \alpha_\mathbb{R}-\alpha_\mathbb{P}$ while $p(0,0) = 2 \alpha_\mathbb{R} - 1 - \alpha_\mathbb{P}$. Inserting these values in (\ref{alphaP}) gives 
\beqn{alphaP1}
 & 0 = 2(\alpha_\mathbb{R}-\alpha_\mathbb{P}) - (2 \alpha_\mathbb{R} - 1 - \alpha_\mathbb{P}) +  C_{RL}(0,0), \quad {\rm i.e.} \\
 &  \alpha_\mathbb{P} =  1+ C_{RL}(0,0) = 1 +  \sum_{m_R, m_L \ge 1}  \dfrac{c(m_R,m_L) }{m_R! m_L!}(-1)^{m_R+m_L} = 1 +\dfrac{ \langle n_R n_L \rangle - \langle n_R \rangle \langle n_L \rangle}{Y} + \dots \nonumber 
\eeqn
The last equation reproduces the claim of \cite{Veneziano:1973nw} yielding 
a unit intercept for the (bare) Pomeron when $R-L$ correlations are negligible (so that the total pressure is the sum of the pressures of the two species according to Dalton's law) \footnote{The prediction $ \alpha_\mathbb{P} = 1$ was also derived by H. Lee \cite{Lee:1972jj} under the stronger assumption of no correlations even within each species.} and extends it to the case in which  they are not.

Comparing the Regge behavior of $\sigma_{excl}$, $\sigma^{pl}_t$ and $\sigma^{cyl}_t$, we can say that $s^{\alpha_\mathbb{R} - \alpha_\mathbb{P}} = \exp(-Y( \alpha_\mathbb{P}  - \alpha_\mathbb{R}))$ is the price
one has to pay for a valence quark (and its corresponding flavor) to be exchanged between the two initial mesons. An additional price $\exp(-Y( 1 - \alpha_\mathbb{R}))$ has to be paid if one exchanges a valence quark both in the initial and in the final state, which is what occurs in the $2 \to 2$ exclusive cross section.

In the following we shall assume $ |C_{RL}(0,0) | \ll |p(1,0)| \sim 1$ which leads directly to $  |\alpha_\mathbb{P} - 1| \ll 1$.  Phenomenologically, the  Pomeron intercept is  $(\alpha_\mathbb{P} - 1) = 0.08 \pm 0.01$ \cite{ParticleDataGroup:2022pth}.  Justifications for  a small value of the $R-L$ correlator can be given using both dynamical and diagrammatic arguments. The short-range correlators within a single species are due, to a large extent, to the fact that the final stable particles very often result from the decay of resonances. However, the most prominent $q\bar{q}$ resonances can only occur in planar channels involving  sets of ``adjacent" (w.r.t. the planar diagram ordering) final particles. No such resonances can appear in ``mixed" channels, i.e. those containing final particles of both species. The (not completely unrelated) diagrammatic argument consists of the observation that final particles belonging to different species are far away in high order Feynman diagrams that dominate the large distance QCD dynamics. Thus, intuitively, their momentum distributions should be weakly correlated. We can rephrase this last observation by saying that the two excited strings resulting from having separated two $q\bar{q}$ pairs in momentum and position phase space should fragment independently. These consideration will also apply to the processes discussed in the forthcoming sections.

\section{Feynman-Wilson gas with baryons}
\label{sec:FWGB}

In this Section we discuss processes with either one or two (anti)baryons in the initial state. For pedagogical reasons it is easier to describe first the process $B\bar{B} \to mesons$. We will then see that quantities related to that process also appear in other contexts, including baryon transport in meson-baryon scattering, baryon transport in $BB$ and $B\bar{B}$ scattering, and $B\bar{B}$ pair creation in generic processes.

\setcounter{equation}{0}
\subsection{FWG for baryon-antibaryon annihilation}
\label{sec:FWGBAB}

We will discuss the physically relevant case with $N_{c}=3$ and 
for now, as explained in the introduction, limit our considerations to diagrams with the simplest topology consistent with the picture of baryons as a Mercedes star with quarks at the ends and the three ($N_{c}=3$) Wilson lines joining in a point (junction) to make a gauge invariant operator. This structure, developed in ref. \cite{Rossi:1977cy} as a natural extension of the notion of topological expansion in the presence of baryons, was corroborated by the results of ref. \cite{Rossi:2016szw}.

In that approximation the FWG for $B\bar{B} \to mesons$ consists of a gas with three distinct species (associated with each ``page" of the baryon ``book" topology), so that, in analogy with (\ref{CylSigma}), we define: 
\beqn{AnnlSigma}
\Sigma_{ann}(z_1,z_2,z_3) = \frac{1}{\sigma_t^{ann}} \sum_{n_1+n_2+n_3 \ge 2} z_1^{n_1} z_2^{n_2} z_3^{n_3} \sigma^{ann}(n_1,n_2,n_3) ~ \Rightarrow ~ \Sigma_{ann}(1,1,1) =1
\eeqn
where $\sigma_t^{ann}$ is the total annihilation cross section. With a straightforward generalization of (\ref{CylSigma1}) we rewrite Equation (\ref{AnnlSigma})  in the form: 
\beqn{AnnSigma1}
&&\Sigma_{ann}(z_1,z_2,z_3) \equiv e^{Y p(z_1,z_2,z_3)} = \nonumber\\
&&=\exp \left(Y \sum_{m_1+m_2+m_3\geq 1} c(m_1,m_2,m_3) \frac{(z_1-1)^{m_1} (z_2-1)^{m_2} (z_3-1)^{m_3} }{m_1! m_2! m_3!}\right) 
\eeqn
and then split $p(z_1,z_2,z_3)$ in terms of correlations with one species, two species and all the  three species:
\beqn{AnnSigma2}
\hspace{-.2cm}&&p(z_1,z_2,z_3) \!=\!  p_1(z_1)\! +\! p_2(z_2)\! +\!  p_3(z_3) \! +\! C_{12}(z_1,z_2)\!+\! C_{13}(z_1,z_3)\!+\! C_{23}(z_2,z_3)\! +\!  C_{123}(z_1,z_2,z_3), \nonumber \\
 \hspace{-.2cm}&& p_i(z_i) = \sum_{m_i \ge 1} c(m_i, 0,0) \frac{(z_i-1)^{m_i}  }{m_i!} \, , \quad i=1,2,3 \nonumber \\
  \hspace{-.2cm}&& C_{12}(z_1,z_2) =  \sum_{m_1, m_2 \ge 1}  c(m_1,m_2,0) \frac{(z_1-1)^{m_1} (z_2-1)^{m_2} }{m_1! m_2!}~;~ {\mbox{+~cyclic}} \nonumber \\
\hspace{-.2cm} && C_{123}(z_1,z_2,z_3) =  \sum_{m_1, m_2,m_3 \ge 1}  c(m_1,m_2,m_3) \frac{(z_1-1)^{m_1} (z_2-1)^{m_2} (z_3-1)^{m_3} }{m_1! m_2! m_3!}
\eeqn
Three equations come out of (\ref{AnnSigma2}) after using $p_i(1) = C_{ij}(1,z_j) = C_{123}(1,z_j,z_k) = 0$, namely
\beqn{AnnSigma3}
&  p(1,1,0) - p_{ann}(0) = 0 ~;~ \nonumber \\
 &0 =  p(1,1,0) + {\rm cycl} - p(1,0,0) - {\rm cycl} + p(0,0,0) - C_{123}(0,0,0)~;~ \nonumber  \\
 & 0 =  p(1,0,0) + {\rm cycl} - 2 p(0,0,0) + \sum_{ij} C_{ij}(0,0) + 2 C_{123}(0,0,0) 
\eeqn
where we have defined  $p_i(0)=p_{ann}(0)$. Equations 
 (\ref{AnnSigma3}) then lead to the following relations among the different intercepts relevant to $B\bar{B}$ annihilation (see Fig. (\ref{f3}) and Table \ref{tab:tab1}):
\beqn{Annrel}
&\alpha_{{\mathbb{J}_0}} - \alpha_{{\mathbb{J}_2}} = -p_{ann}(0) ~;~ \nonumber  \\
 &\alpha_{{\mathbb{J}_0}} - (2\alpha_{\mathbb{B}}-1)= 3(\alpha_{{\mathbb{J}_2}} - \alpha_{{\mathbb{J}_4}}) -  C_{3}(0,0,0) ~;~ \nonumber  \\
 &(\alpha_{{\mathbb{J}_0}} - \alpha_{{\mathbb{J}_4}}) = 2(\alpha_{{\mathbb{J}_0}} - \alpha_{{\mathbb{J}_2}}) -  C_{2}(0,0)
\eeqn
These relations, illustrated in Fig. (\ref{f4}), show 
 how the equal splitting of the intercepts gets broken by the two and three-species correlations. However, without further input, they simply give the three intercepts $\alpha_{{\mathbb{J}_0}}, \alpha_{{\mathbb{J}_2}},\alpha_{{\mathbb{J}_4}}$ in terms of the (known) value  $\alpha_{{\mathbb{B}}} \sim 0$ and the three unknown parameters $p_{ann}(0),C_{ij}(0,0)$ and $C_{123}(0,0,0)$. In the following and in the figures we will simplify notations and write $C_{ij}(0,0)\equiv C_2, C_3(0,0,0)\equiv C_3$.
\begin{figure}[t]
\centering
\includegraphics[scale=0.28]{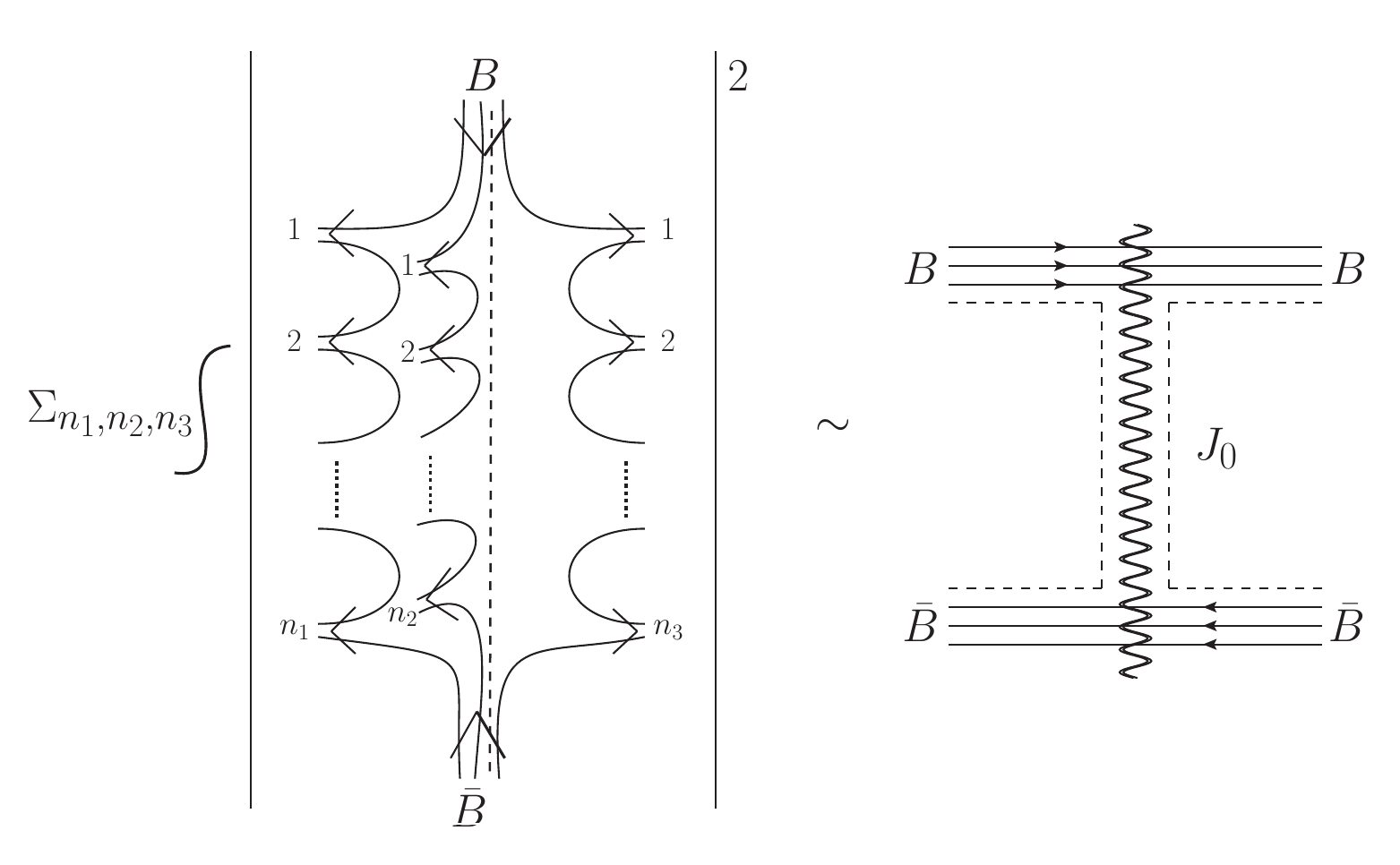}\hspace{.1cm}
\includegraphics[scale=0.28]{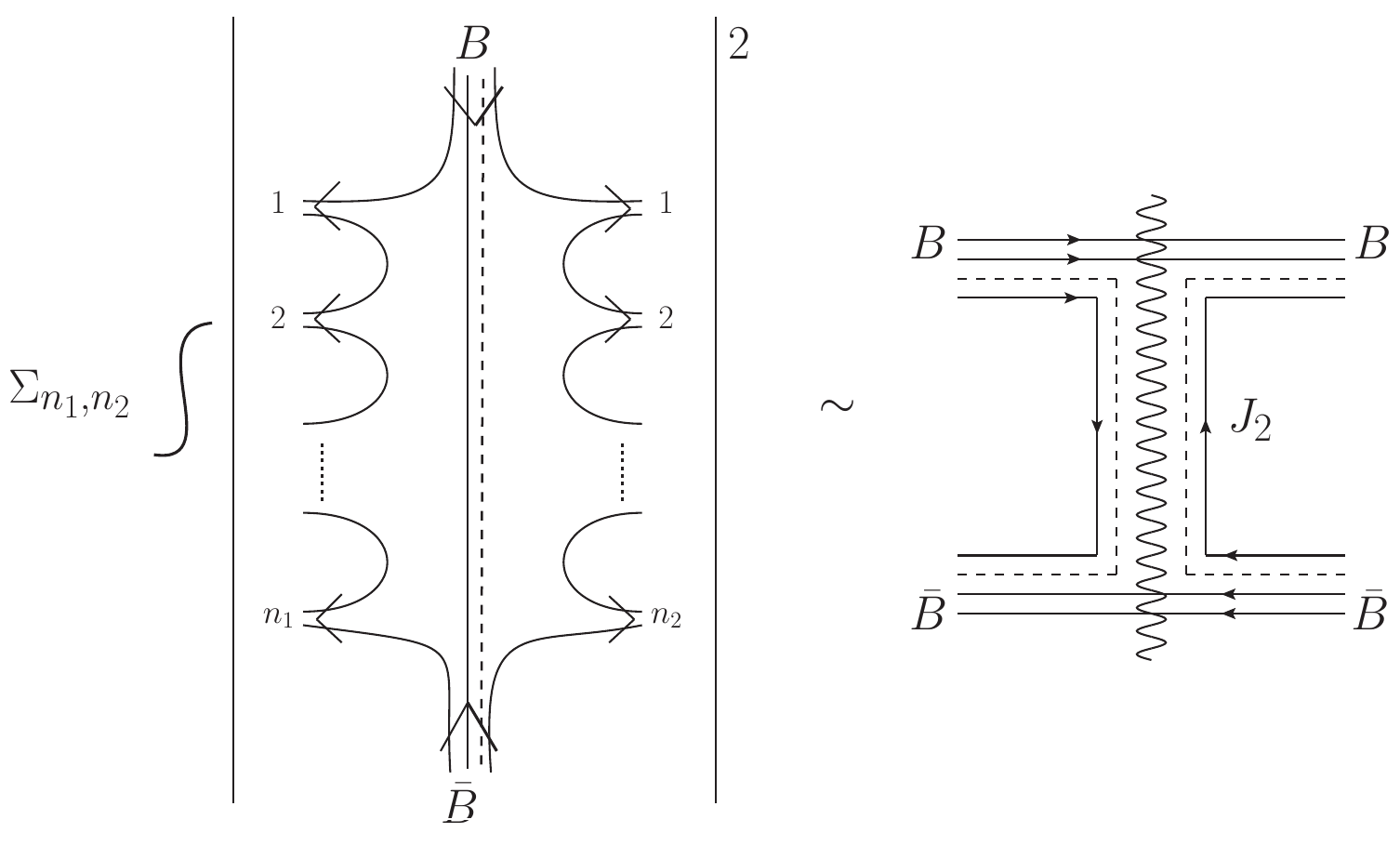}
\caption{First two pictures from the left: total $B\bar{B}$ annihilation cross section and the Regge-pole $J_0$ controlling its high-energy limit. Last two pictures: $B\bar{B}$ annihilation cross section with a $s$-channel $q\bar{q}$ annihilation and the Regge-pole $J_2$ controlling its high-energy limit.}
\label{f3}
\end{figure}

\begin{figure}[t]
\centering
\includegraphics[width=\textwidth]{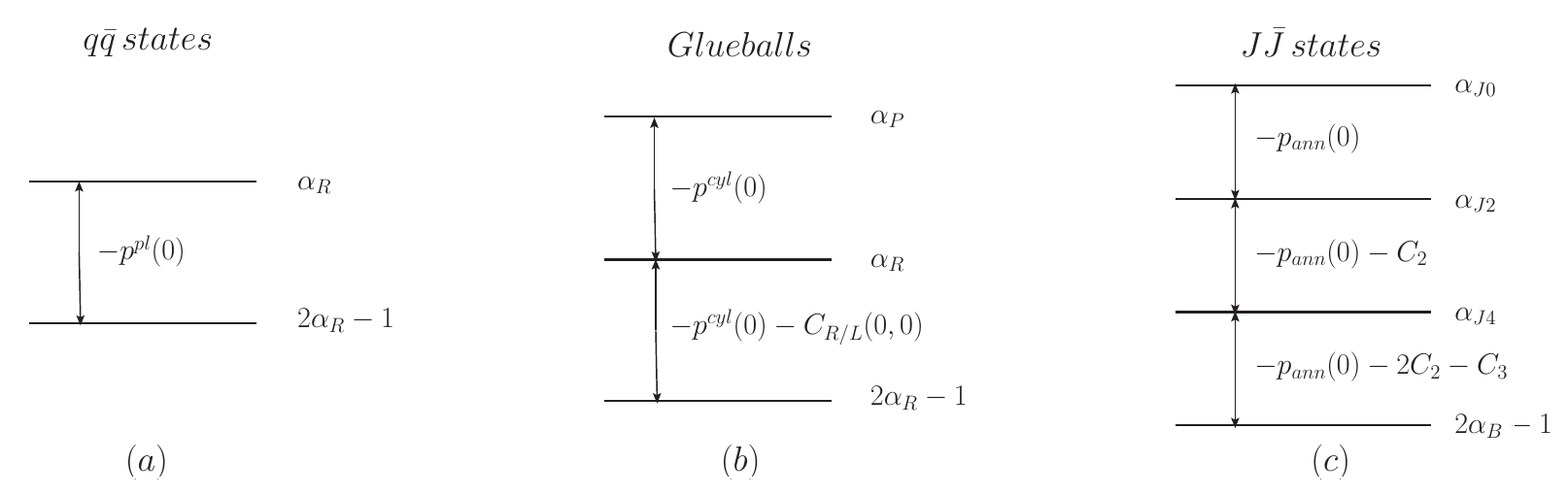}
\caption{Intercepts of the Regge poles controlling the different cross sections discussed in Sections \ref{sec:PFWG} and \ref{sec:FWGC}. The same  intercepts will appear in the rapidity dependence of flavor and baryon-number transport discussed in Section \ref{sec:TFBN}. In this figure for uniformity we have introduced the following notations. We have denoted by $p^{pl}(0)$ the ``planar'' pressure $p(0)$ defined in Equation (\ref{plSigma1}), by $p^{cyl}(0)$ the ``cylinder'' pressure $p_L(0)=p_R(0)$ defined in Equation (\ref{CylSigma2}) and by $p_{ann}(0)$ the pressure $p_i(0)$ defined in Equation (\ref{AnnSigma2}).}
\label{f4}
\end{figure}

An interesting open question is whether one can relate the jumps in pressure at the cylinder or planar level  to those encountered in $B\bar{B}$ annihilation. In the case in which inter-species correlations are neglected it is reasonable to assume that they are all the same and this leads to the predictions made in \cite{Rossi:1977cy}. However, in the presence of inter-species correlations there is some ambiguity on how to connect the pressures in different processes.

One possibility is to identify $p_{ann}(0)$ in (\ref{AnnSigma3}) with the individual pressure at the cylinder level, $p^{cyl}(0) = \alpha_\mathbb{R} - \alpha_\mathbb{P}$, defined in Fig. \ref{f4}, and $C_{2}$ of  (\ref{AnnSigma2}) with $C_{RL}(0,0)$ in (\ref{CylSigma2}) i.e. with $(\alpha_\mathbb{P} -1)$. In this case we predict: 
\beqn{Annrel0}
 &\alpha_{{\mathbb{J}_4}} = (2\alpha_{\mathbb{B}}-1) +  (1 -  \alpha_\mathbb{P}) +(1- \alpha_\mathbb{R}) -  C_3 \sim - 0.5 -  C_{RL} - C_3   \nonumber
  \\
 &\alpha_{{\mathbb{J}_2}}= (2\alpha_{\mathbb{B}}-1) +  (1 -  \alpha_\mathbb{P}) + 2(1- \alpha_\mathbb{R}) -  C_3 \sim 0 - C_{RL} - C_3 \nonumber \\
 &\alpha_{{\mathbb{J}_0}}= (2\alpha_{\mathbb{B}}-1)  + 3 (1- \alpha_\mathbb{R})  -  C_3 \sim 0.5 - C_3 \eeqn
Interestingly, with this identification the value of $\alpha_{\mathbb{J}_0}$ turns out to be the same as the original one \cite{Rossi:1977cy} up to the three-species correlation $C_3$. It is reasonable to expect that such high-order correlation is even smaller than $C_{RL}\sim (\alpha_\mathbb{P} - 1)$.

Another possibility is to identify $p_{ann}(0)$ in (\ref{AnnSigma3}) with the pressure at the planar level, $p(0)$ in (\ref{alphaR}) i.e.\ with $\alpha_\mathbb{R}-1 \sim -0.5$  while still associating $C_{2}$ with $(\alpha_\mathbb{P}-1)$. In this case we would predict instead of Eq.(\ref{Annrel0}): 
\beqn{Annrel1}
 &\alpha_{{\mathbb{J}_4}} = (2\alpha_{\mathbb{B}}-1) +  2(1 -  \alpha_\mathbb{P}) +(1- \alpha_\mathbb{R}) -  C_3 \sim - 0.5 - 2 C_{RL} - C_3   \nonumber
  \\
 &\alpha_{{\mathbb{J}_2}}= (2\alpha_{\mathbb{B}}-1) +  3(1 -  \alpha_\mathbb{P}) + 2(1- \alpha_\mathbb{R}) -  C_3 \sim 0 - 3 C_{RL} - C_3 \nonumber \\
 &\alpha_{{\mathbb{J}_0}}= (2\alpha_{\mathbb{B}}-1) + 3(1 -  \alpha_\mathbb{P}) + 3 (1- \alpha_\mathbb{R})  -  C_3 \sim 0.5 - 3 C_{RL} - C_3 \eeqn
 The data favors this option; if we take $C_{RL} = 0.08$, and $C_3=0$, we reproduce the data. Indeed, we get $\alpha_{{\mathbb{J}_0}}\simeq 0.26$ yielding the beam rapidity dependence $e^{(\alpha_{\mathbb{J}_0} + \alpha_\mathbb{P} - 2)Y/2} = e^{-0.66\ Y/2}$ in Eq.\ (\ref{eq:rapidity_dist}), to be compared to the STAR experimental result $e^{-0.65\ Y/2}$ \cite{Lewis:2022arg}. 

Note that if the $C_{RL}$ and $  C_3$ correlations were negligible, the two predictions outlined above in Eq.(\ref{Annrel0}) and Eq.(\ref{Annrel1}) would agree and give back the original estimates of \cite{Rossi:1977cy}.

\section{Regge trajectories and spectra of the junction-antijunction glueballs} \label{sec:glueball}
\setcounter{equation}{0}

Here we will try to relate the $J_0$ trajectory to the spectroscopy of potential junction-antijunction glueballs. We will assume the usual linear Regge trajectory
\be{eq:traj}
\alpha(M^2) = \alpha(0) + \alpha' M^2,
\ee
relating the spin and mass of hadrons on the trajectory. Thus, the intercept $\alpha(0)$ can be extracted if one knows the spin and mass of any particle on the trajectory, as well as the slope $\alpha'$. 

The QCD operator corresponding to the junction-antijunction $J_0$ glueball is  given by a contraction of three Wilson lines \cite{Rossi:1977cy} (see Eq. (\ref{WL})):
\beq
\hspace{-.3cm}G_{\JJ}(x,y)\! =\! \epsilon^{ijk}\epsilon_{i'j'k'} [U(C_1(x,y))]_i^{i'}
[U(C_2(x,y))]_j^{j'}
[U(C_3(x,y))]_k^{k'},
\label{GJJN}
\ee
where the integrations are along three different paths, $C_1, C_2$ and $C_3$.  Note that this operator does {\it not} go into $\pm$ itself under charge conjugation $\cal C$. Indeed it goes to a similar operator where junction and antijunction are exchanged and, at the same time, the directions of the three Wilson lines are reversed. On the other hand it is clear that the new state has the same mass spectrum as the original one because QCD respects $\cal C$-invariance. \footnote{Under $\cal C $ the states $J_2$ and $J_4$ get transformed into their antiparticles.} The above argument suggests  that the $J_0$ Regge pole actually corresponds to two degenerate $\cal C$-related Regge poles of opposite signature, denoted by $f_0^J$ and $\omega_0^J$ in Ref. \cite{Rossi:1977cy}. Their degeneracy ensures that they combine to give a purely real amplitude in $BB$ scattering corresponding to the absence of an annihilation channel there. Obviously this degeneracy will be lifted if the two $\cal C$-eigenstates mix with other states with the same $\cal C$, such as $\mathbb{P}$ for $f_0^J$ and $\omega$ for $\omega_0^J$. However, this mixing should be suppressed by the JOZI rule (see \cite{Montanet:1980te} for a discussion of the JOZI rule). 

Let us see how the $\cal C$ parity of the $J_0$ states can be described in a perturbative language. We will start by performing the expansion of the operator (\ref{GJJN}) up to the 2-gluon order, $\cO(g^2)$. With each Wilson line expanded to 0, 1 or 2 orders the following contributions are present in the expansion:
\begin{itemize}
\item $\cO(g)\times \cO(g)\times \cO(1) : G_{gg1} = g^2\,\tr\left(\int_{C_1}A \int_{C_2}A +  \text{2 permutations}\right)$ 
\item $\cO(g^2)\times \cO(1)\times \cO(1) : G_{g^211} = -g^2\,\tr\left(P(\int_{C_1}A)^2+  \text{2 permutations}\right )$.
\end{itemize}

Here we used the shorthand notation: $$\int_{C_1}A\equiv\int_{C_1(x,y)} dz^\mu A_\mu(z); ~~~ P\left(\int_{C_1} A\right)^2 \equiv \int_{C_1(x,y)} dz^\mu A_\mu(z) \int_{C_1(x,z)} dz'^\nu A_\nu(z').$$ 

Under the $\cC$-conjugation the gluon field (which is a SU(3) matrix) transforms like  $A\rightarrow -A^T$ as can be seen from its coupling to the quark fields in the QCD Lagrangian \footnote{Since $\cC$-conjugation acts on the fermion field as $C\psi C = -i(\bar{\psi}\gamma^0\gamma^2)^T, C\bar{\psi}C = -i(\gamma^0\gamma^2\psi)^T$, the transformation $C A_{\mu j}^i C = - A_{\mu i}^j$ ensures that the quark-gluon interaction term in the QCD Lagrangian is invariant under $\cC$: $$C\bar{\psi}_i\gamma^\mu A_{\mu j}^i \psi^j C = A_{\mu i}^j (\gamma^0\gamma^2\psi^i)^T\gamma^\mu(\bar{\psi}_j\gamma^0\gamma^2)^T = - A_{\mu i}^j  \bar{\psi}_j\gamma^0\gamma^2(\gamma^\mu)^T\gamma^0\gamma^2\psi^i = A_{\mu i}^j \bar{\psi}_j\gamma^\mu \psi^i$$}. Therefore, both terms in the 2-gluon expansion are ${\cal{C}}$-even.

Let us now discuss the properties of the corresponding Regge trajectories. The 2-gluon state in the $S$-wave (corresponding to the lowest-lying state on the  Regge trajectory) is $\mathcal{P}$-even. Since we are interested in the leading $J_0$ Regge trajectory, we need to identify the $S$-wave two gluon state with the highest spin. 
This is a spin 2 state.  

\begin{figure}
    \centering
\includegraphics[scale=.3]{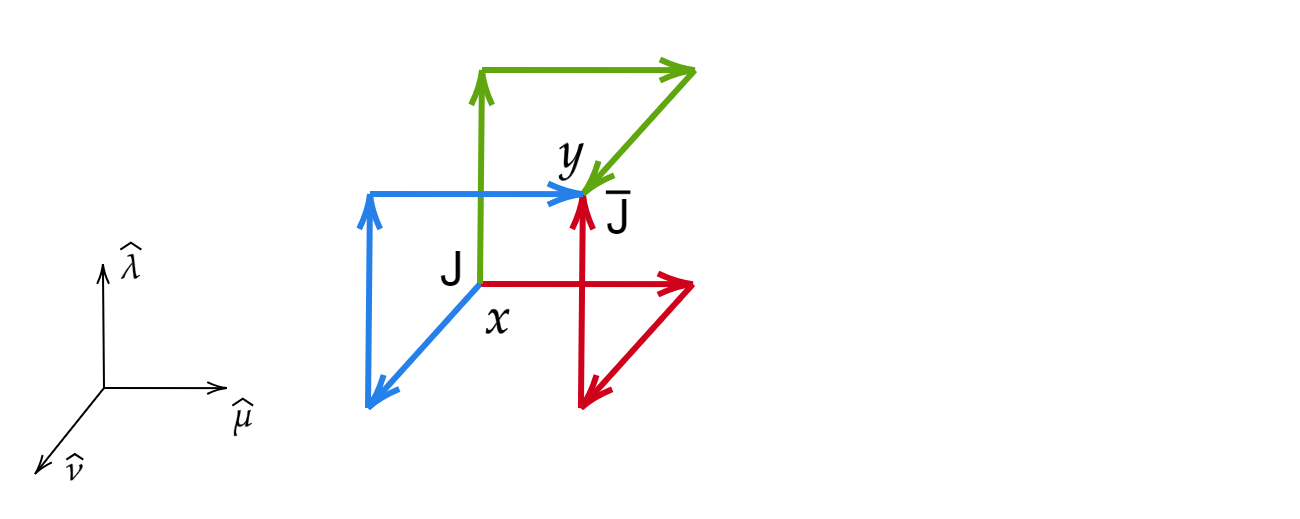}
    \caption{A possible lattice operator that can be used to search for the $J_0$ glueball. The color is antisymmetrized at each of the junctions. Directions on the lattice are introduced in accordance with Eq. (\ref{eq:J0_lattice}).}
    \label{fig:JJ_lattice}
\end{figure}

A glueball state with  $J^{\cal{PC}} = 2^{++}$ was reported in a quenched lattice calculation \cite{Chen:2005mg}. Even though  it has the right quantum numbers, we do not expect it to be a $J_0$ glueball since the operator used in the computation did not include the junction-antijunction color structure. We thus propose to perform a lattice calculation of the $J_0$ masses using, for instance, the operator (\ref{GJJN}) with the three paths as depicted in Fig. (\ref{fig:JJ_lattice}):
\begin{align}
&& O_{J_0}=\epsilon_{i_1 j_1 k_1}U^{i_1}_{i_2}(x,\hat{\mu})U^{i_2}_{i_3}(x+\hat{\mu},\hat{\nu})U^{i_3}_{i_4}(x+\hat{\mu}+\hat{\nu},\hat{\lambda})U^{j_1}_{j_2}(x,\hat{\nu})U^{j_2}_{j_3}(x+\hat{\nu},\hat{\lambda}) \nonumber 
\\
&& U^{j_3}_{j_4}(x+\hat{\nu}+\hat{\lambda},\hat{\mu})U^{k_1}_{k_2}(x,\hat{\lambda})U^{k_2}_{k_3}(x+\hat{\lambda},\hat{\mu})U^{k_3}_{k_4}(x+\hat{\lambda}+\hat{\mu},\hat{\nu}) \epsilon^{i_4 j_4 k_4}, \label{eq:J0_lattice} \end{align}
where $U^i_j(x,\hat{\mu})$ denotes a gauge link operator originating at point $x$ in the direction $\hat{\mu}$, see Fig.\ \ref{fig:JJ_lattice} for notations.

To predict the mass of the ${J^{\cal{PC}}}=2^{++}$ $J_0$ glueballs, we need to know the slope of the $J_0$ Regge trajectory. One can relate this slope to the slope of the Pomeron trajectory. Since the Pomeron exchange corresponds to a closed string, we can estimate the $J_0$ trajectory slope to be 
\beq
\alpha_{{\mathbb{J}_0}}' \simeq \frac{2}{3} \alpha_\mathbb{P}' \simeq 0.10 - 0.17 \,\text{GeV}^{-2}, \label{eq:J0_slope}
\ee
where the numerical value is obtained by using the Pomeron slope in the range  $\alpha_\mathbb{P}' \simeq 0.15 - 0.25\,\text{GeV}^{-2}$ corresponding to various estimates \cite{ParticleDataGroup:2022pth, ParticleDataGroup:2014cgo}.\ Using the intercept of the leading $J_0$ Regge trajectory $\alpha_{{\mathbb{J}_0}}\simeq 0.25 - 0.5$, we obtain the prediction for the ${J^{\cal{PC}}}=2^{++}$ $J_0$ glueball mass in the range of $M_{J_0} = 3.0 - 4.1$ GeV. These values of the mass are highlighted in Fig. (\ref{fig:traj}) together with the Regge trajectories constraining the mass range as discussed above.

\begin{figure}
    \centering
\includegraphics[width=0.6\linewidth]{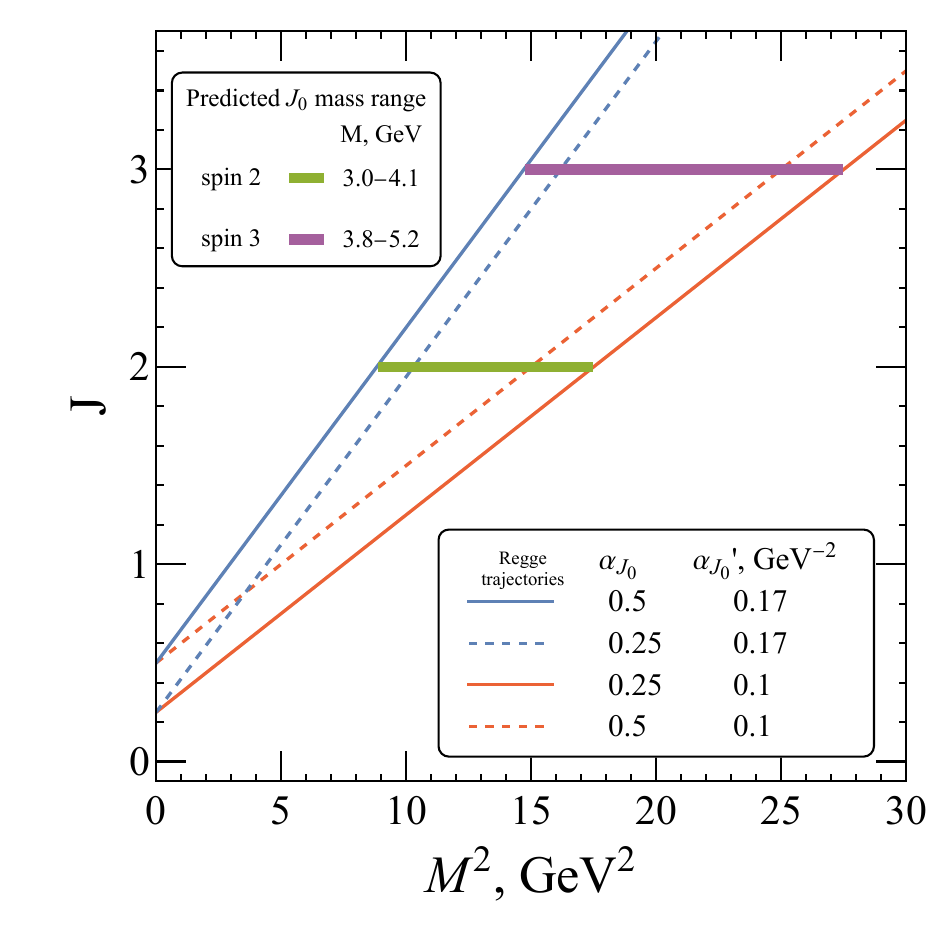}
    \caption{Leading $J_0$ Regge trajectories corresponding to the various values of the intercept. The predicted mass range of spin-two and spin-three glueballs is highlighted.}
    \label{fig:traj}
\end{figure}

It is also instructive to repeat this analysis in the case of the 3-gluon terms in the expansion of (\ref{GJJN}). One obtains the following $\cO(g^3)$ terms: 
\begin{itemize}
\item $\cO(g)\times \cO(g)\times \cO(g) : G_{ggg} = -i\,g^3\, \tr \left(\int_{C_1}A \int_{C_2}A \int_{C_3}A + \int_{C_1}A \int_{C_3}A \int_{C_2}A \right)$,
\item $\cO(g^2)\times \cO(g)\times \cO(1) : G_{g^2g1} = \dfrac{i\,g^3}{2} \tr \left(P(\int_{C_1}A)^2 \int_{C_2}A + \text{5 permutations}\right)$,
\item $\cO(g^3)\times \cO(1)\times \cO(1) : G_{g^311} = -\dfrac{i\,g^3}{3} \tr \left(P(\int_{C_1}A)^3 +  P(\int_{C_2}A)^3 + P(\int_{C_3}A)^3\right)$,
\end{itemize}
where the following additional notation was introduced:
$$P\left(\int_{C_1} A\right)^3 \equiv \int_{C_1(x,y)} dz^\mu A_\mu(z) \int_{C_1(x,z)} dz'^\nu A_\nu(z')  \int_{C_1(x,z')} dz''^\lambda A_\lambda(z'').$$
Along the lines of the previous ${\cal C}$-parity discussion, the $G_{ggg}$ term is $\cC$-odd. Indeed, it is easy to see that the operator $G_{ggg}$ includes only terms containing the symmetric structure constants, $d_{abc} \propto \tr(\lambda^a \lambda^b \lambda^c + \lambda^a \lambda^c \lambda^b)$. Such symmetric 3-gluon combinations are indeed $\mathcal{C}$-odd, see e.g.\ \cite{Novikov:1977dq}.\
On the other hand, $G_{g^2g1}$ and $G_{g^3 11}$ are a mixture of $\cC$-even and $\cC$-odd operators because under $\cC$-conjugation $P(\int_{C_1} A)^n\rightarrow (-1)^n \bar{P}(\int_{C_1} A)^n$, where $\bar{P}$ denotes inverse path-ordering. 

The 3-gluon state in the $S$-wave is $\mathcal{P}$-odd, therefore the highest-spin 3-gluon component of the $J_0$ can be $J^{\cal{PC}} = 3^{--}$ and $J^{\cal{PC}} = 3^{-+}$.
Estimate for the masses of these glueballs similar to those provided by Eq. (\ref{eq:J0_slope}) yield values that fall in the range of $3.8 - 5.2$\, GeV as displayed in Fig. \ref{fig:traj}.

The large uncertainty in the range of $J_0$ glueball masses thus predicted is mainly due to the uncertainty of the Pomeron trajectory slope (\ref{eq:J0_slope}) that we use to estimate the slope of the $J_0$ trajectory. In Section \ref{sec:exp} we suggest a way to extract the slope of the $J_0$ trajectory directly from doubly-diffractive baryon-antibaryon pair creation in $pp$ or $ep$ collisions. Such a measurement would allow to further narrow down the range of the glueball masses that we expect to be found in the lattice simulations.

\section{Transport of flavor and baryon number}
\label{sec:TFBN}
\setcounter{equation}{0}

In this section we extend the arguments of the previous sections to the study of inclusive spectra as a function of rapidity (i.e.\ after integration over the transverse momenta).
The main point we wish to make is that, in the topological expansion approach, flavor and baryon number transport over large rapidity intervals are very weakly correlated. Furthermore, the exponents that control the exponential decay of the various inclusive cross sections as a function of rapidity intervals are not independent and the relations between them can be predicted theoretically and checked against the data.

The tools that we will employ in this part of the paper are those discussed in the Sections \ref{sec:PFWG}, \ref{sec:FWGC} and \ref{sec:FWGB} up to two important extensions:
\begin{itemize}
\item We shall need the more  detailed description of the inclusive cross sections. This could in principle include information about the transverse momenta of the detected final particles but for our purposes it will be sufficient to discuss distributions in rapidity $y$ i.e.\ after integration over the (assumed strongly cut-off) transverse momenta, see Appendix A for details. We will therefore deal with single particle distribution of the type:
\beqn{flt}
 \rho(y_f) =\frac{1}{\sigma^{cyl}_t}  \frac{d \sigma^{cyl}(1+2 \ra y_f + X)}{d y_f}
\eeqn
and with their generalization to the multiparticle distributions defined in  Appendix~A.
\item We will also use Mueller and Kancheli's  extension \cite{Mueller:1970fa, Kancheli:1970gt} of the optical theorem and Regge pole analysis to evaluate the inclusive cross sections.
\end{itemize}
\noindent We stress again that, as in the previous Sections, all this will be done in the context of the TE of QCD~\cite{Veneziano:1976wm}.

\subsection{Quark/flavor transport in meson-meson scattering}

Let us first consider, at the bare-Pomeron level, the rapidity dependence of quark transport from the fragmentation regions $y \sim \pm Y/2$ to some $y$ such that $|\pm Y/2 - y| \gg 1$. If the valence quark to be transported is the one initially at $y \sim Y/2$, one can analyze the single particle distribution (\ref{flt})
using the Mueller-Kancheli generalized optical theorem \cite{Mueller:1970fa, Kancheli:1970gt}.

 The Mueller-Kancheli diagram describing this inclusive cross section is dominated by the leading $q\bar{q}$ Regge trajectory from $y= Y/2$ down to $y = y_f$ and by the bare-Pomeron pole from $y = y_f$ to $y = -Y/2$ (see right panel in Fig. (\ref{f5})). 
 In drawing the Mueller-Kancheli diagram we  use the convention of showing in each $t=0$ channel, the quark and or junction lines that flow through the corresponding rapidity gap.
 
For instance, in Fig. (\ref{f5}) we are considering the regime:
$|\pm Y/2 - y_f| \gg 1$. If the valence quark to be transported is  initially at $y \sim Y/2$ then the leading trajectory in the interval  
$[Y/2 - y_f]$ contains a $q\bar{q}$ pair whereas the trajectory determining the leading behaviour at large $[y_f +Y/2]$ is the vacuum (Pomeron) trajectory. As a result one predicts  the single particle distribution (\ref{flt}) to behave as 
\beqn{flt1}
 & \rho(y_f) \sim \exp(-Y  \alpha_\mathbb{P}) \exp( \Delta y ~\alpha_\mathbb{R}) \exp((Y- \Delta y)  \alpha_\mathbb{P} ) \nonumber \\
 &= \exp(- \Delta y ( \alpha_\mathbb{P}-\alpha_\mathbb{R}))~;~\quad \Delta y  \equiv (Y/2 - y_f)
\eeqn
where the first exponential comes from the normalization factor $\sigma^{cyl}_t$ in the definition (\ref{flt}).
\begin{figure}[t]
\centering
\includegraphics[scale=0.4]{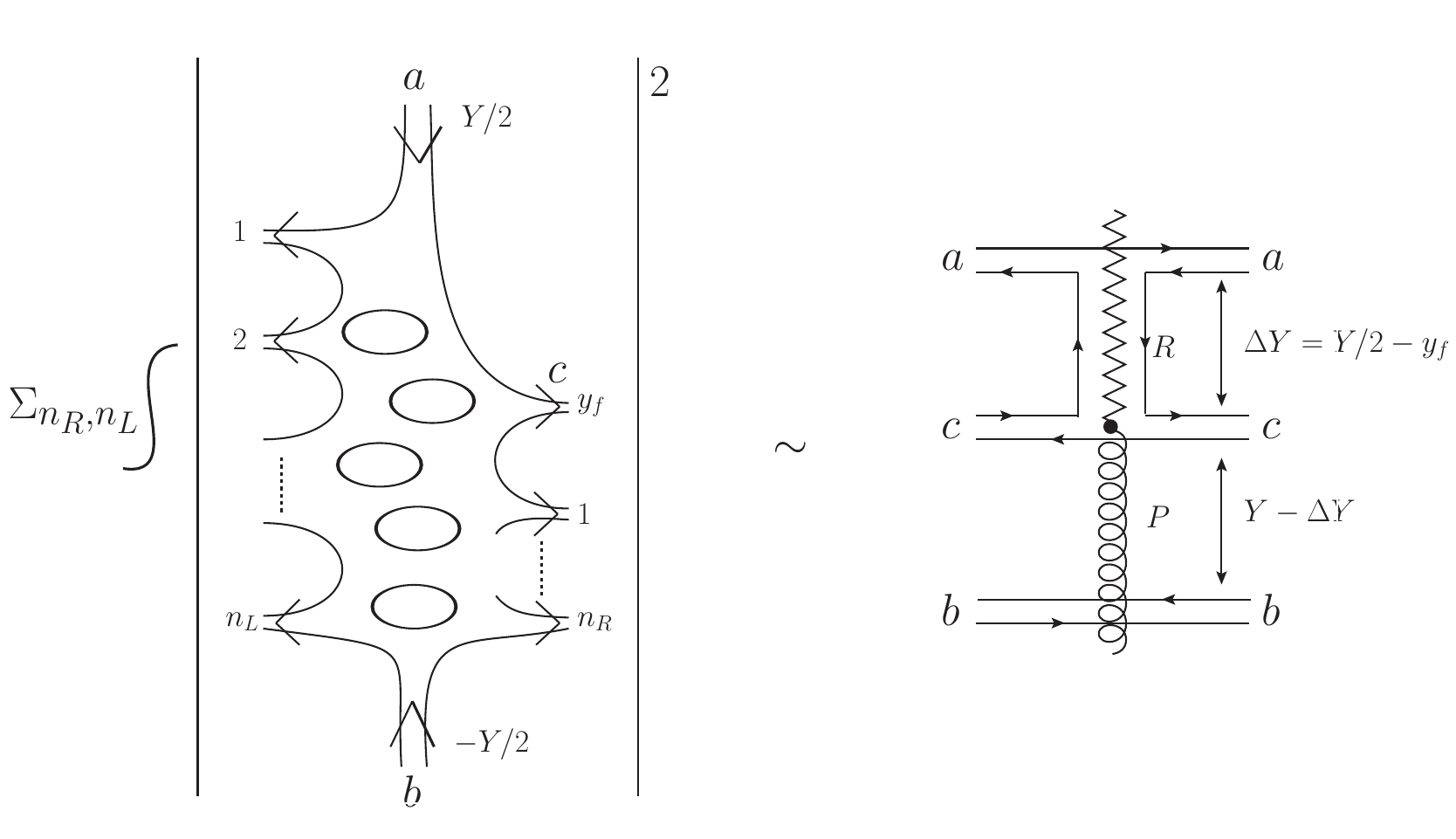}
\caption{Flavor transport in rapidity in $MM$ scattering and its corresponding Mueller-Kancheli diagram leading to (\ref{flt1}).}
\label{f5}
\end{figure}

In analogy with what we discussed at the end of Section \ref{sec:PI} we see that the price to be paid for transporting a quark from $y = Y/2$ to $y = y_f$ is given by (recall the second line of eq. (\ref{alphaP1}))
 \beqn{price}
  \exp(- |Y/2 - y_f| ( \alpha_\mathbb{P}-\alpha_\mathbb{R})) = \exp(- |Y/2 - y_f| (1 + C_{RL}(0,0)-\alpha_\mathbb{R})) 
\eeqn 
In the liming case in which $y_f = -Y/2$ we recover the penalty $ \exp(- Y( \alpha_\mathbb{P}-\alpha_\mathbb{R}))$ discussed in Section \ref{sec:PI} for annihilating a quark between the two initial mesons.
Furthermore, one expects quark transport to be accompanied by a reduced multiplicity (by about a factor 2) in the rapidity interval $[Y/2 - y_f]$ compared to the complementary interval $[y_f - (-Y/2 )]$, because in one case we cut a Reggeon/plane  and in the other a Pomeron/cylinder.

It is interesting  theoretically to discuss the 
total transfer in rapidity of a given conserved flavor $Q^{(i)}$ ($i = u, d, s, \dots$) carried by the initial mesons. In order to compute such a quantity we would have to sum the contribution to $Q^{(i)}$ coming from each individual  meson's inclusive cross section. This is quite a  hard task that we leave to further investigations. However, we can use some sort of completeness argument (often used in jet-physics or heavy quark fragmentation) to connect directly the rapidity distribution of the valence quark in the final state as tracing the total transfer of the particular charge it carries. The idea is that somehow that charge will have to be deposited in {\it some} meson sitting at the quark's rapidity.

With this proviso, we can readily give the $(Y, y_f)$ dependence of the charge distribution in the final state in a form similar to the one given in Eq. (\ref{eq:rapidity_dist}) for the baryon-number distribution:
 \beqn{Qdistr}
\frac{d Q^{(i)}}{d y_f} \propto e^{(\alpha_i + \alpha_{\mathbb{P}}-2)Y/2}[ g_{ia} e^{(\alpha_{\mathbb{P}}-\alpha_i)y_f} + g_{ib} e^{(\alpha_i -\alpha_{\mathbb{P}})y_f} ],
\eeqn
where $\alpha_i$ stands for (the intercept of) the leading $q_i \bar{q_i}$ trajectory   ($\alpha_i = \alpha_{\rho}$ for $i = u,d$; $\alpha_i = \alpha_{\phi}$ for $i = s$, \dots) and $g_{ia}, g_{ib}$ are the couplings of that trajectory to the two initial particles. Note that, unlike the case of the inclusive single particle cross section (\ref{flt1}), we are not normalizing the distribution to the total cylinder cross section.

One can consider linear combinations of the above expressions in order to describe, for instance, the transport of electric-charge $Q \sim \frac23 Q^{(u)} - \frac13 Q^{(d)} - \frac13 Q^{(s)} $ that can be directly measured in experiment. On the other hand is is easy to check that no net charge transport occurs for the combination $B \equiv \frac13\sum_i Q^{(i)}  $corresponding to baryon number. For this to happen one needs at least one baryon in the initial state as discussed in the following sub-sections.

\subsection{Baryon and flavor transport in meson-baryon scattering}

The case of meson-baryon scattering is very similar to the one of meson-meson scattering as far as flavor transport is concerned. In particular, Eqs. ((\ref{flt1})), (\ref{Qdistr})  are still valid. The process is analogous to the one shown in Fig. (\ref{f5}) with particle $a$ being  now a baryon.

However, a new feature now appears, namely baryon-number transport. In other words, if we fix the initial baryon's rapidity to be $+Y/2$, we can ask what is the inclusive rapidity distribution of a final baryon sitting at some $y_f$. 
Proceeding in analogy with our treatment of conserved-charge transfer in meson-meson scattering we may first consider the  inclusive cross section for a given final baryon and get, in analogy with ((\ref{flt1})),
\beqn{MB}
& \rho_B(y_f) \sim \exp[-|Y/2- y_f| ( \alpha_\mathbb{P}- \alpha_{{\mathbb{J}_0}})] \sim \exp(-0.82\ |Y/2- y_f| ),
\eeqn
where we have used the value of $\alpha_{{\mathbb{J}_0}}$ from Eq.(\ref{Annrel1}). The corresponding Mueller-Kancheli diagram is shown in Fig.\ (\ref{f6}). 

\begin{figure}[t]
\centering
\includegraphics[scale=0.4]{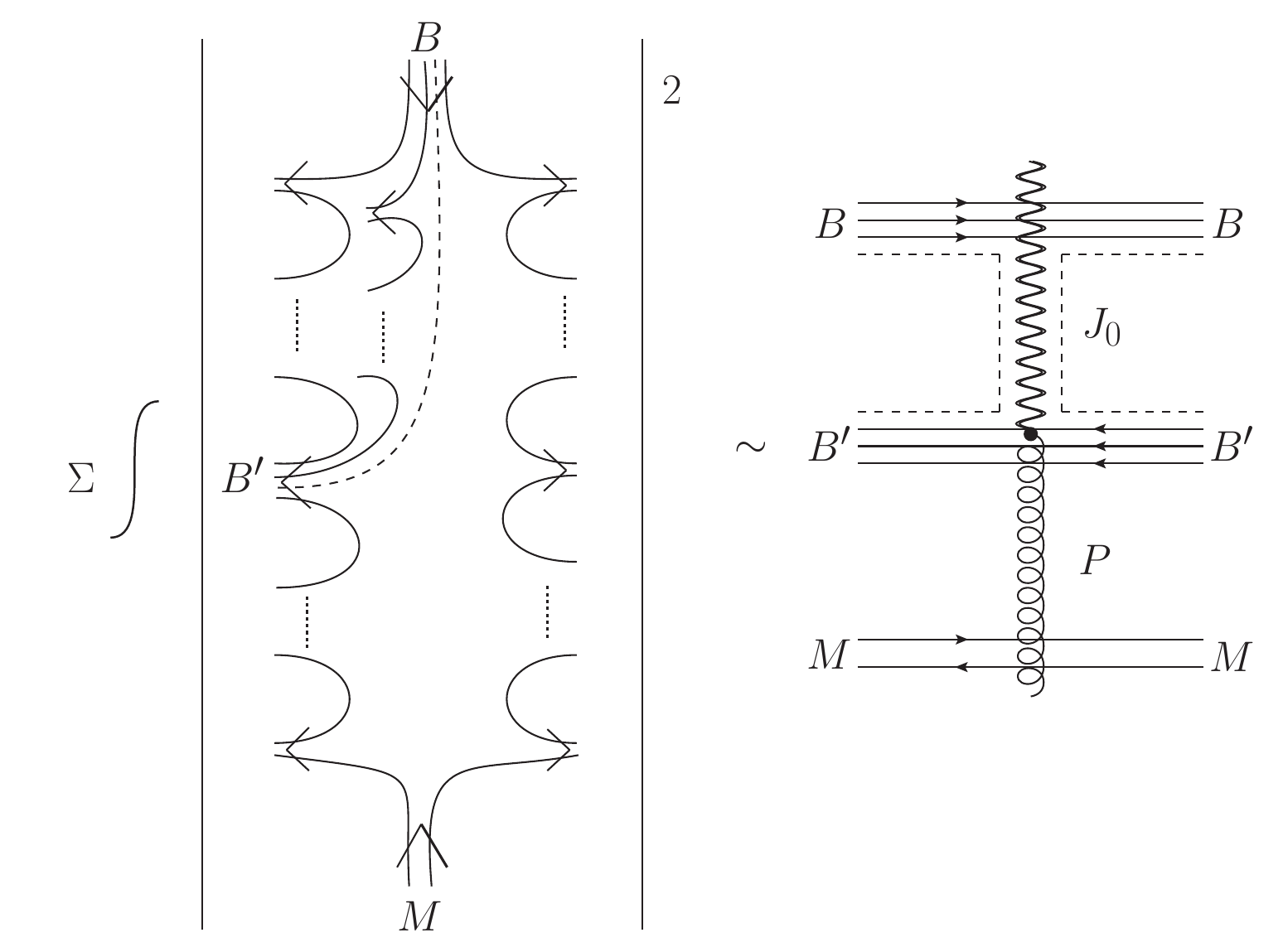}
\caption{Baryon-number (without valence quarks) transport  in $MB$ scattering and its corresponding Mueller-Kancheli diagram leading to equation (\ref{MB}).}
\label{f6}
\end{figure}

As in the case of flavor transport in meson-meson scattering we can consider the ``super-inclusive" baryon-number distribution, i.e. summed over the individual single baryons. The resulting distribution (not normalized to the total cross section) will have the following rapidity dependence:
\beq
\frac{d N}{d y_f} \propto e^{(\alpha_{\mathbb{J}_0} + \alpha_{\mathbb{P}}-2)Y/2} e^{(\alpha_{\mathbb{P}}-\alpha_{\mathbb{J}_0})y_f} , 
\label{eq:MB_dist}
\eeq
with much similarity with Eq. (\ref{eq:rapidity_dist}). 

Quite exceptionally,  one or more of the initial quarks 
may end up  in the final baryon.  The corresponding rapidity distribution would be dominated by the $J_2$ and $J_4$  trajectories according to
\begin{align}
 \rho_B(y_f) \sim  \exp[-|Y/2-y_f|(\alpha_\mathbb{P} - \alpha_{\mathbb{J}_2})], \nonumber \\
 \rho_B(y_f) \sim  \exp[-|Y/2-y_f|(\alpha_\mathbb{P} - \alpha_{\mathbb{J}_4})],
\label{eq:MB_subleading}
\end{align}
for one and two transported quarks, respectively. We postpone the discussion of  joint baryon-number-charge transfer to the case of baryon-baryon scattering presented in the following subsection.  

\subsection{Baryon and flavor transport in baryon-(anti)baryon scattering}

We now move to baryon-(anti)baryon scattering which, together with its extension to heavy-ion collisions, is of course the most interesting case. The difference with respect to the $MB$ case discussed in the previous subsection is that we now  have either two junctions or a junction-antijunction pair in the initial state. We have already discussed, in Section (\ref{sec:FWGBAB}), the case of $B\bar{B}$ (i.e.\ junction-antijunction) annihilation. Here we will consider instead the case in which the baryon-number connected with an initial junction (or antijunction) is transported towards small rapidities. We will do so while still neglecting baryon-antibaryon pair creation, a topic discussed at the end of this Section.

As discussed in the previous subsection, the leading contribution will come from  $J_0$ exchange: 
\beqn{BB}
 & \rho_B(y_f) \sim \exp[-|Y/2- y_f| ( \alpha_\mathbb{P}- \alpha_{{\mathbb{J}_0}})] \sim \exp(-0.82 |Y/2- y_f| ) . 
\eeqn

\begin{figure}[t]
\centering
\includegraphics[scale=0.27]{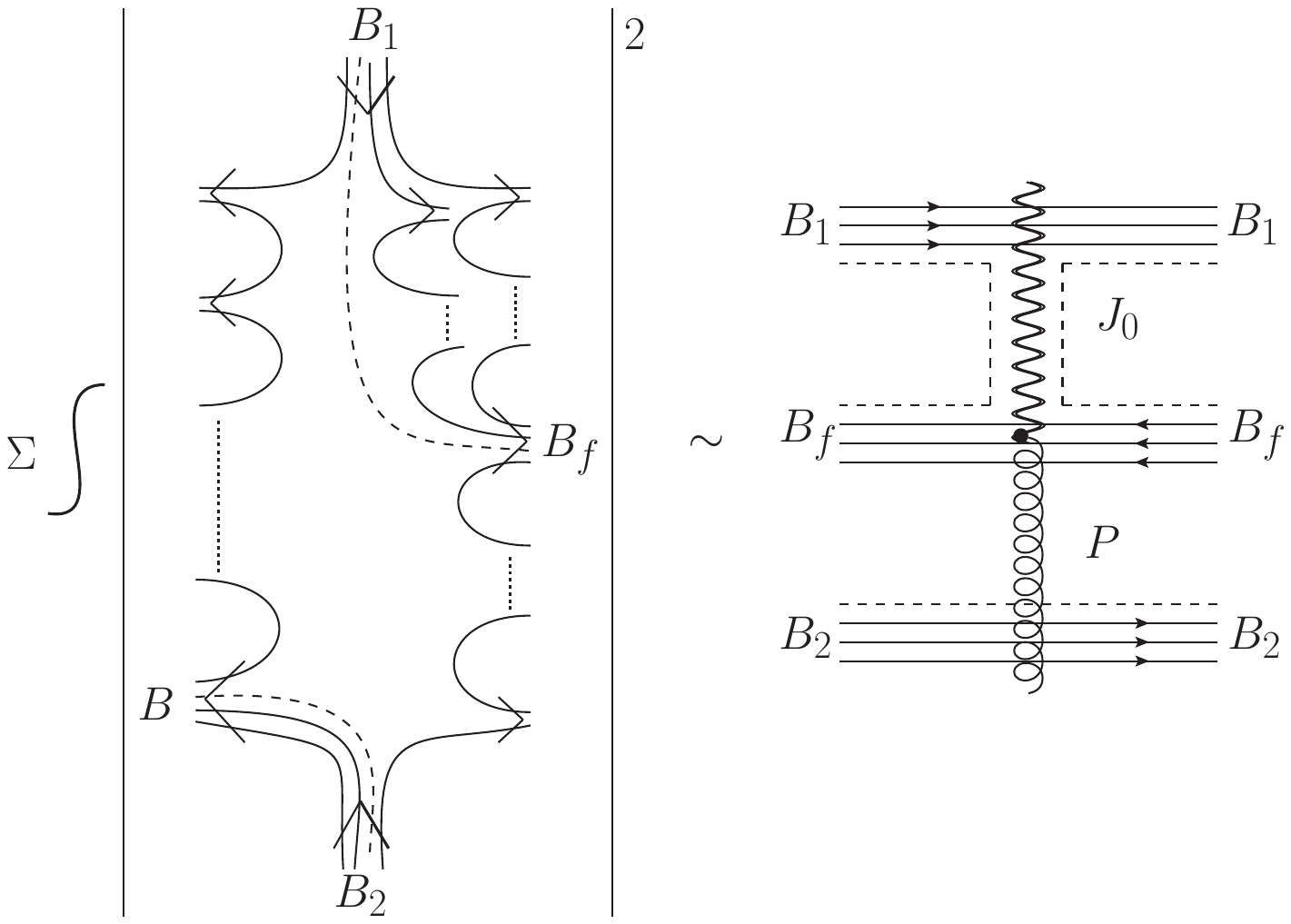}
\includegraphics[scale=0.27]{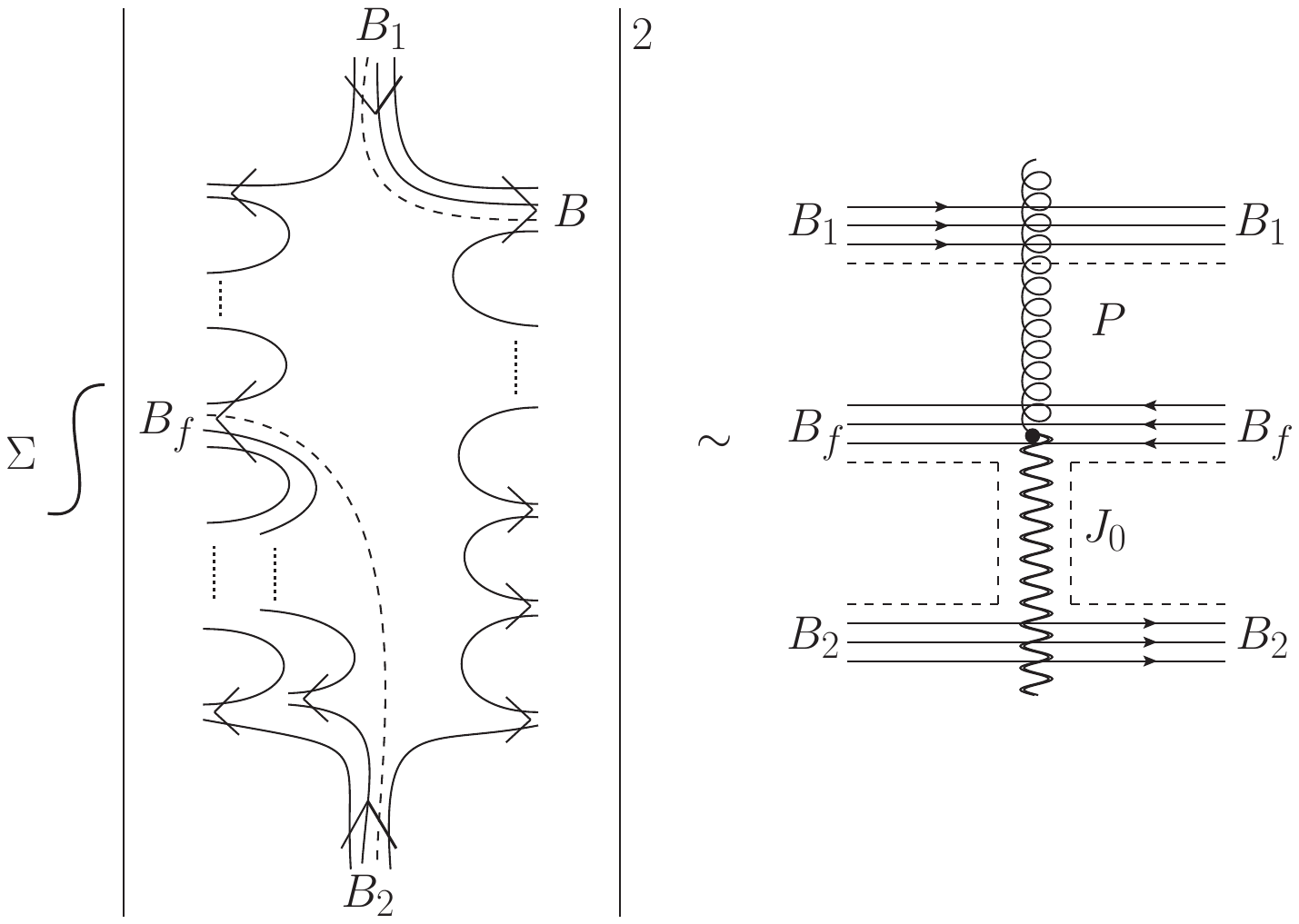}
\caption{Left panel: baryon-number (without flavor) transport in $BB$ scattering via the junction of $B_1$ and its corresponding Mueller-Kancheli diagram. Right panel: the same but with baryon-number transport via the junction of $B_2$. The two contributions are added
up in (\ref{BB1}).}
\label{f8ab}
\end{figure}

Actually, the only difference with respect to $MB$  scattering is that now the transported baryon number can come from either one of the two initial baryons. Therefore, a term with $Y/2 \ra -Y/2$ has to be added to (\ref{BB}).  The single-baryon inclusive cross section is thus 
\beqn{BB1}
 \hspace{-.4cm}\rho_B(y_f) \sim e^{-(\frac{Y}{2}- y_f) ( \alpha_\mathbb{P}- \alpha_{{\mathbb{J}_0}})} + e^{-(\frac{Y}{2}+ y_f) ( \alpha_\mathbb{P}- \alpha_{{\mathbb{J}_0}}) } =  2 e^{-\frac{Y}{2} ( \alpha_\mathbb{P}- \alpha_{{\mathbb{J}_0}})} \cosh[y ( \alpha_\mathbb{P}- \alpha_{{\mathbb{J}_0}})]
\eeqn
This situation is illustrated in Fig. (\ref{f8ab}). Once more it makes more sense, experimentally, to trace baryon-number transport ``super-inclusively" i.e. without reference to the specific baryon detected in the final state. Arguing as in the cases of meson-meson and meson-baryon scattering, Eq. (\ref{BB1}) leads precisely to  Eq. (\ref{eq:rapidity_dist}). 

One may also consider the (unlikely) case in which the observed final baryon transports to rapidity $y_f$ both the junction and one quark of $B_1$.
This will give  
\beqn{BB2}
 \rho_B(y_f) \sim  e^{-(\frac{Y}{2} - y_f) ( \alpha_\mathbb{P}- \alpha_{{\mathbb{J}_2}})}
 \eeqn
as illustrated in Fig. (\ref{f8abc}) (left panel),
 \begin{figure}[t]
\centering
\includegraphics[scale=0.28]{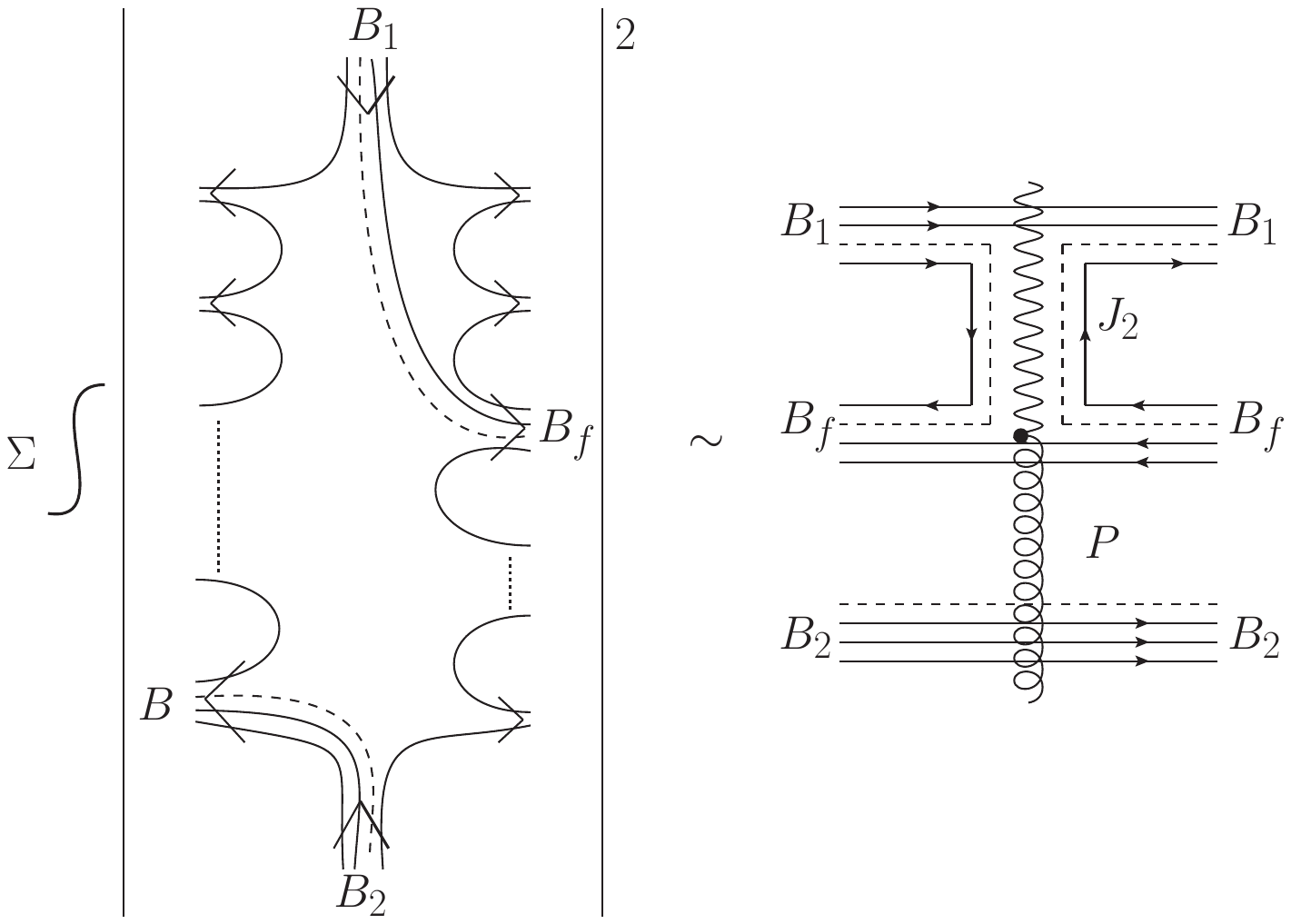}
\includegraphics[scale=0.28]{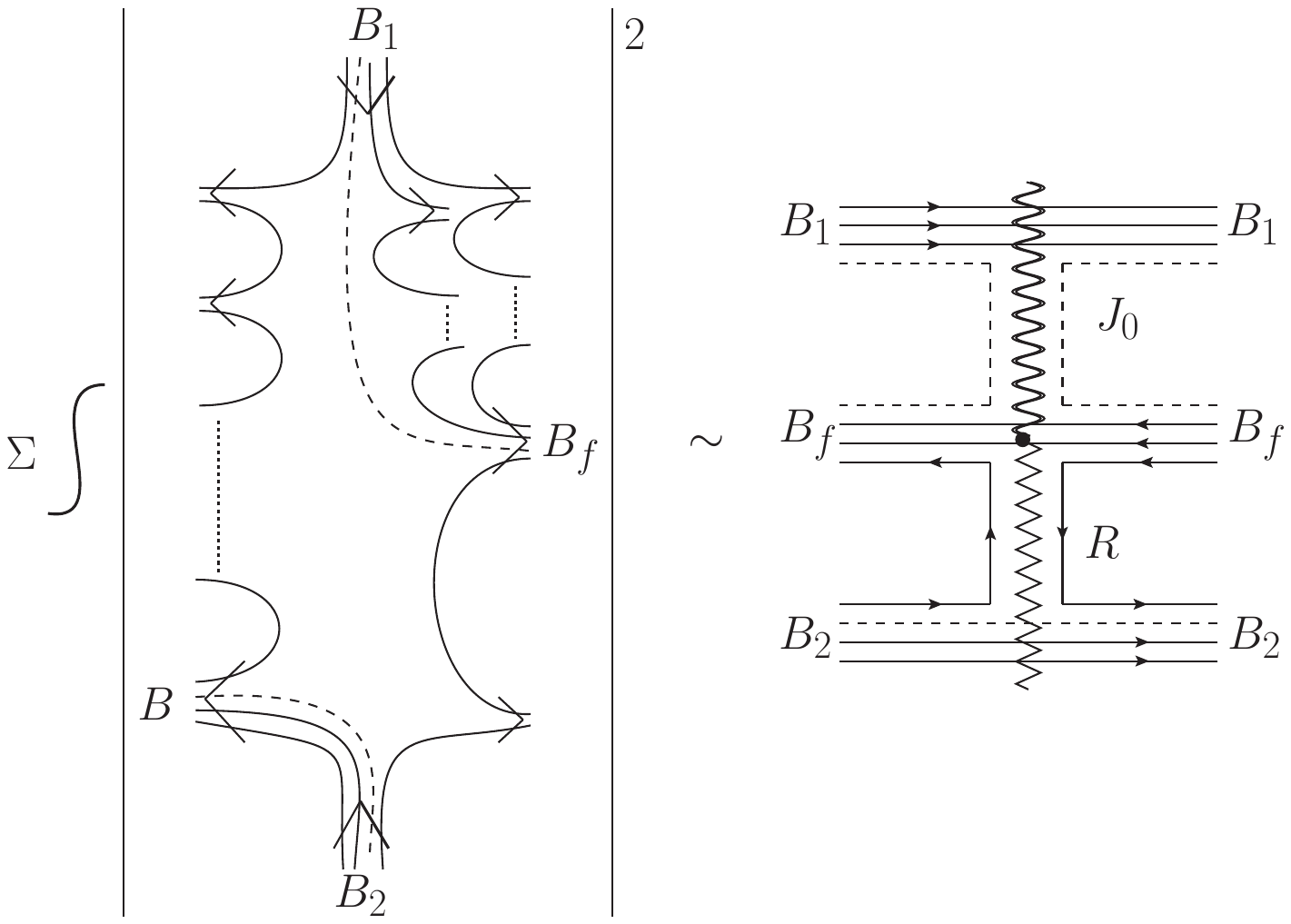}
\caption{Left panel: baryon-number plus one-quark transport from $B_1$ and its corresponding Mueller-Kancheli diagram leading to (\ref{BB2}). Right panel: baryon-number transport from $B_1$ together with a one-quark-transport from $B_2$ and its corresponding Mueller-Kancheli diagram leading to eq. (\ref{BB3}).}
\label{f8abc}
\end{figure}
while the case in which the observed baryon transports the junction of $B_1$ and one valence quark of $B_2$, yields 
\beqn{BB3}
 \rho_B(y_f) \sim  e^{-(\frac{Y}{2} - y_f) ( \alpha_\mathbb{P}- \alpha_{{\mathbb{J}_0}})}e^{-(\frac{Y}{2} + y_f) ( \alpha_\mathbb{P}- \alpha_\mathbb{R})}
 \eeqn
This case is illustrated in Fig. (\ref{f8abc}) (right panel). These processes are suppressed with respect to $J_0$ exchange (no valence quark transport), described by Eq.(\ref{BB1}). 

\subsection{Combined charge--baryon-number rapidity distribution}

Let us finally come to the very interesting topic of the combined baryon-number/ charge transport in baryon-baryon collisions, a quantity which is expected to neatly differentiate between the junction and the valence-quark pictures for the tracing of baryon number. To this purpose we have to consider a two-particle inclusive cross section or, even better, a joint distribution of charge and baryon number. In the junction picture it is unlikely for one of the initial junctions and one of the initial valence quarks to end up in the same baryon sitting at small $y_f$. Much more frequently, the valence quark will end up in a meson. Consider then the combined baryon-number--charge 
distribution:
 \beqn{BFdistr}
 \frac{d^2\sigma_{(B,Q)} }{dy_B dy_Q} \equiv F_{(B,Q)}(Y, y_B, y_Q)\, ,
 \eeqn
 as an extension of the separate flavor and baryon-number distributions discussed so far.
 
With arguments basically identical to those used for each separate distributions it is straightforward to discuss the dependence of $F_{(B,Q)}$ both on the total rapidity $Y$ and on the rapidity differences $|(\pm \frac{Y}{2} - y_B)|, |(\pm \frac{Y}{2} - y_Q)|, | y_B-  y_Q|$ all taken to be sufficiently large. The number of distinct cases is a bit larger than in the single-distribution case, but still quite manageable. Up to a trivial exchange of the two incoming baryons, we can identify four distinct cases whose Mueller-Kancheli diagrams are given in Figs. (\ref{f11a}) and (\ref{f11b}). The corresponding distributions are given by:
\begin{align}
F^{(1)}_{(B,Q)} &\propto e^{(\alpha_\mathbb{P} + \alpha_{\mathbb{J}_2}-2)\frac{Y}{2}}~e^{(\alpha_R-\alpha_{\mathbb{J}_2})y_B} ~ e^{(\alpha_{\mathbb{P}}-\alpha_R)y_Q} \; , \nonumber \\
F^{(2)}_{(B,Q)} &\propto  e^{( \alpha_{\mathbb{P}} + \alpha_{\mathbb{J}_2}  -2) \frac{Y}{2}}~ e^{(\alpha_{\mathbb{J}_0} -\alpha_{\mathbb{J}_2}) y_Q} ~ e^{(\alpha_{\mathbb{P}}-\alpha_{\mathbb{J}_0}) y_B} \; , \nonumber \\
F^{(3)}_{(B,Q)} &\propto  e^{( \alpha_R+ \alpha_{\mathbb{J}_0}  -2) \frac{Y}{2}}~ e^{(\alpha_{\mathbb{P}}-\alpha_R) y_Q} ~
 e^{(\alpha_{\mathbb{J}_0} - \alpha_{\mathbb{P}}) y_B} \, ,  \nonumber \\
  F^{(4)}_{(B,Q)} &\propto  e^{( \alpha_R+ \alpha_{\mathbb{J}_0}  -2) \frac{Y}{2}}~ e^{(\alpha_{\mathbb{J}_0} -\alpha_{\mathbb{J}_2}) y_Q} ~
 e^{(\alpha_{\mathbb{J}_2} - \alpha_R) y_B} \, . 
\label{eq:BQdistr}
\end{align}
 \begin{figure}[t]
\centering
\includegraphics[scale=0.4]{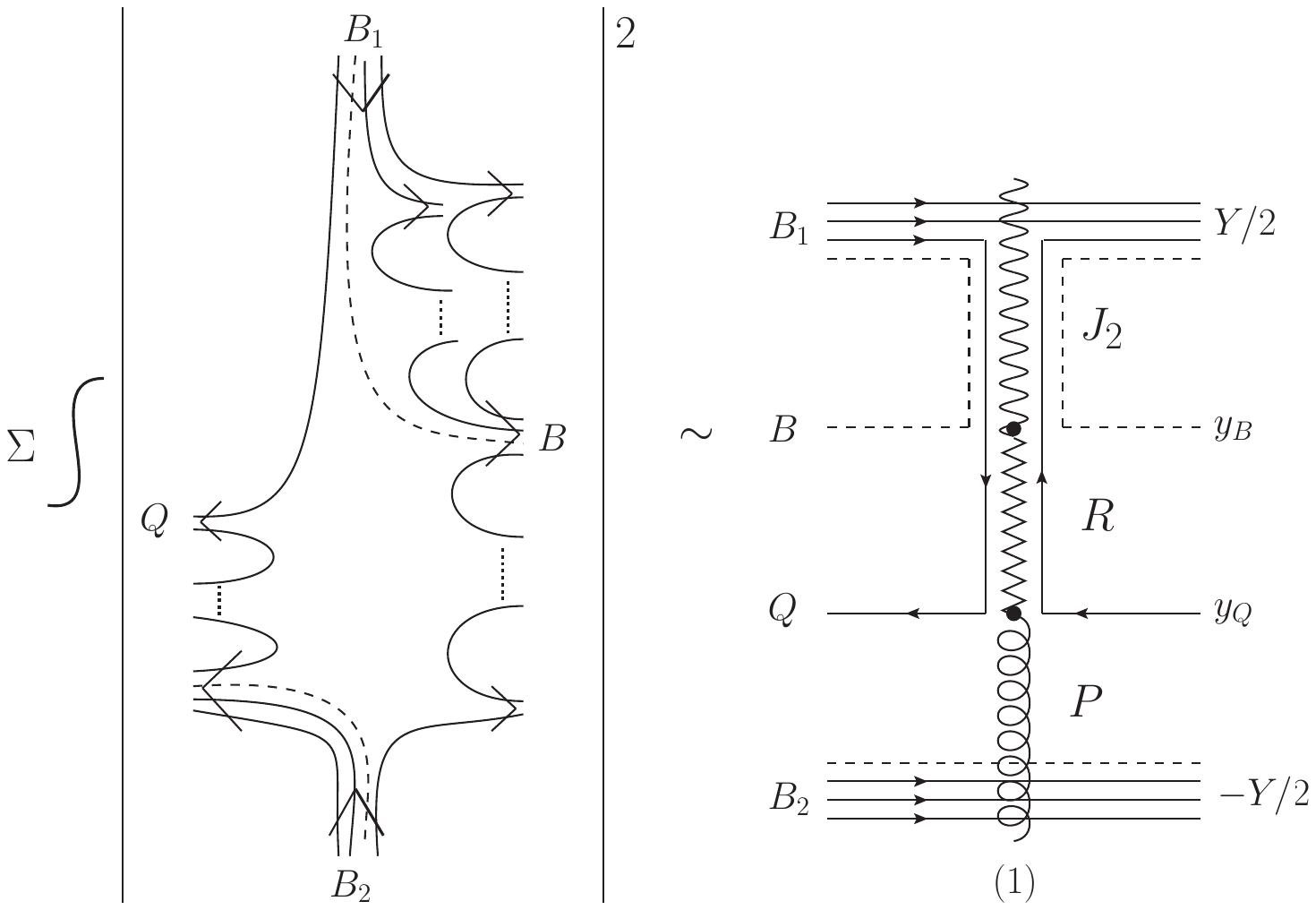}
\caption{Diagrams contributing to charge--baryon-number separation and the corresponding Mueller-Kancheli diagram leading to the first of Eqs. (\ref{eq:BQdistr}). The single quark and junction lines stand for a sum over all the hadrons containing those lines.}
\label{f11a}
\end{figure}
 \begin{figure}[t]
\centering
\includegraphics[scale=0.4]{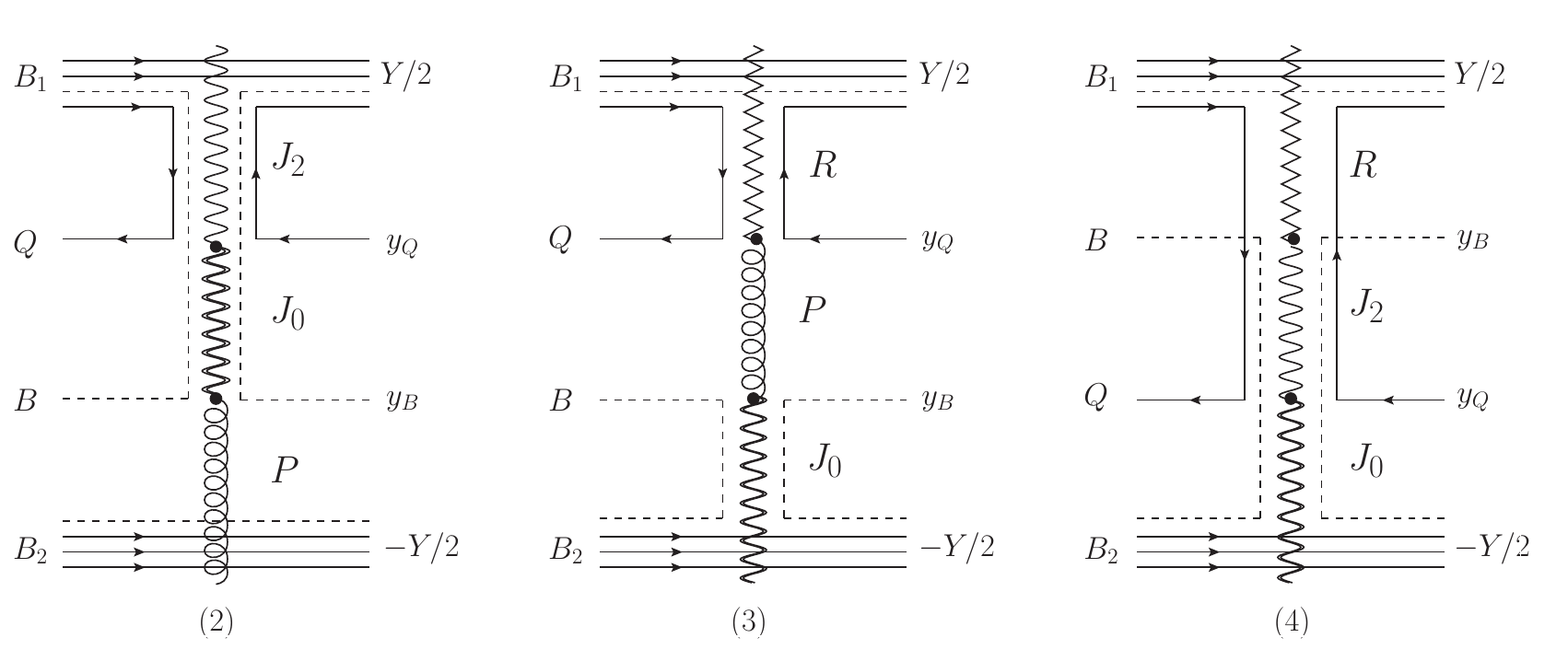}
\caption{The Mueller-Kancheli diagrams leading, in the order, to the last three Eqs.(\ref{eq:BQdistr}). The single quark and junction lines stand for a sum over all the hadrons containing those lines.}
\label{f11b}
\end{figure}
As one can see from Figs. (\ref{f11a}) and (\ref{f11b}), the terms $F^{(1)}_{(B,Q)}$ and $F^{(4)}_{(B,Q)}$ contribute in the case of the rapidity ordering $-Y/2 < y_Q < y_B < Y/2$ while $F^{(2)}_{(B,Q)}$ and $F^{(3)}_{(B,Q)}$ correspond to the ordering $-Y/2 < y_B < y_Q < Y/2$. Of course, for a given $y_B, y_Q$ all relevant contributions (including those related by some trivial symmetry  to the ones of Figs.(\ref{f11a}) and (\ref{f11b})) have to be included. Note that both $J_0$ and $J_2$ play an important role in the above formulae. Therefore, if it is possible to measure those double rapidity distributions with sufficient accuracy, one can extract not only the leading intercept $\alpha_{\mathbb{J}_0}$, but also  $\alpha_{\mathbb{J}_2}$ and thus check the consistency of the framework.

Eq.(\ref{eq:BQdistr}) can be normalized by the total cross section $\sigma_t^{cyl} \sim e^{(\alpha_\mathbb{P} - 1)Y}$ to yield the combined flavor-baryon probability distributions of a kind similar to Eq.(\ref{flt}):
\beq
\rho_{(B,Q)}^{(i)} \propto e^{(1-\alpha_\mathbb{P})Y} F^{(i)}_{(B,Q)}.
\eeq
Finally, we can consider the baryon-flavor correlation function:
\beq
\label{FBcorr}
C_{(B,Q)}(Y, y_B, y_Q) ) \equiv \frac{\rho_{(B,Q)}(Y, y_B, y_Q)}{\rho_B(y_B)~ \rho_Q(y_Q)} -1 \, ,
\eeq
 study its dependence upon the flavor-baryon separation $\Delta y \equiv|y_B-y_Q|$, and contrast it with expectations in the valence-quark picture.
 We have checked that this correlation quickly vanishes as one increases $\Delta y $ while keeping $|y_B+y_Q|$ small, but leave a detailed study of this important topic to future work.

\subsection{Rapidity distribution of $B\bar{B}$-pair creation }

We may finally turn to a process we have so far ignored (since it should only provide small corrections for the cases we have previously considered) i.e.\ $B\bar{B}$-pair production. This process can be thought of as a process $M^* M^* \to B\bar{B} + mesons$  in which the initial mesons are off-shell and spacelike.
The process occurs through the creation of a junction-antijunction pair (presumably from a purely gluonic process). As long as the pair  is not much separated in rapidity, this is a low-energy non-perturbative process we cannot say much about. 

However, as one takes the pair to be well separated in rapidity, the two-particle (meaning  $B\bar{B}$) distribution will fall off exponentially in  $|\Delta y|$. The corresponding exponent is controlled again by the intercept of $J\bar{J}$ trajectories. If we do not veto meson production in the rapidity interval $\Delta y$, 
$\alpha_{{\mathbb{J}_0}}$ will provide the leading contribution for large $|\Delta y|$, giving
\beqn{pair}
 \rho_{B,\bar{B}}(\Delta y) \sim  e^{-  |\Delta y| ( \alpha_\mathbb{P}- \alpha_{{\mathbb{J}_0}})}\, ,
 \eeqn
with subleading contributions coming from $\alpha_{{\mathbb{J}_2}}$, $\alpha_{{\mathbb{J}_2}}$.\ If, instead, meson production is completely vetoed, the 2-baryon cut intercept $(2\alpha_{\mathbb{B}}-1)$ will replace $\alpha_{{\mathbb{J}_0}}$ in (\ref{pair}). The case corresponding to (\ref{pair}) is illustrated in Fig. (\ref{f10}). We stress again that no new input is needed in order to make these predictions. The universality and factorization of Regge poles ensures that the same trajectories control the asymptotic behavior of different measurable processes.

 \begin{figure}[t]
\centering
\includegraphics[scale=0.4]{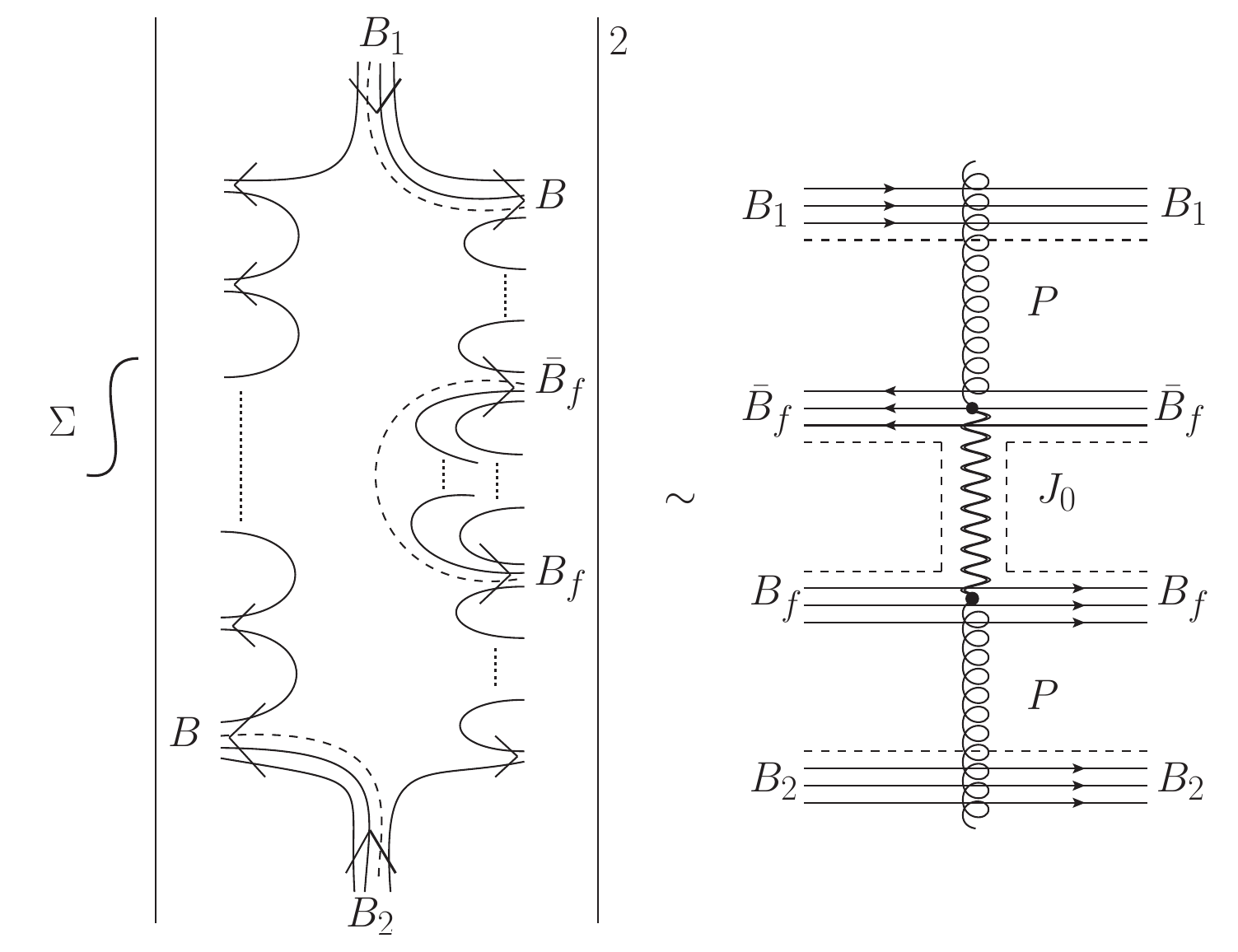}
\caption{$B-\bar{B}$ pair creation and the corresponding Mueller-Kancheli diagram leading to (\ref{pair}).}
\label{f10}
\end{figure}

Let us now come back to the question of the role of $B\bar{B}$-pair production in determining the physical Pomeron intercept discussed in Section \ref{sec:FWGC} where $B\bar{B}$-pair production was not included (although it is certainly in the data). It is quite obvious from Fig. (\ref{f10}) that from a Mueller-Kancheli  point of view adding $B\bar{B}$-pair production will amount to introducing a $\mathbb{P}-{\mathbb{J}_0}$ coupling and mixing which will eventually renormalize upwards the Pomeron intercept. This is allowed if we remember that $J_0$ has a positive $\cal C$, positive signature component which will mix with the Pomeron and thus contribute equally to $BB$ and $B\bar{B}$ total cross sections. However, this mixing should be suppressed by the JOZI rule.

\section{Predictions and experimental tests}
\label{sec:exp}
\setcounter{equation}{0}

Our suggestions and predictions for experiment and lattice calculations can be summarized as follows.
\begin{itemize}
\item The cleanest example of baryon-number -- flavor separation is the stopping of $\Omega$ hyperons in $pp$ or $ep$ collisions. Observing it would represent a clear signature of the baryon-number -- flavor separation phenomenon.

\item At a more inclusive and quantitative level one should develop further the idea of computing and measuring the joined charge-baryon-number rapidity distribution as a function of their rapidity separation. This where the junction picture should drastically differ from the conventional one associating the flow of baryon number to that of the valence quarks.

\item One needs to confirm the energy and rapidity dependence of single baryon stopping, and extract the precise value of the slope of net baryon distribution in rapidity, at a fixed energy of the $pp$ collison, or in semi-inclusive deep-inelastic scattering \cite{Frenklakh:2023pwy}. 

\item We suggest to measure the energy and rapidity dependences of double baryon stopping in $pp$ collisions. We expect that if the stopped baryons are close in rapidity, the rapidity distribution of the pair is flat \cite{Kharzeev:1996sq}.

\begin{figure}
\centering
\includegraphics[width=0.6\linewidth]{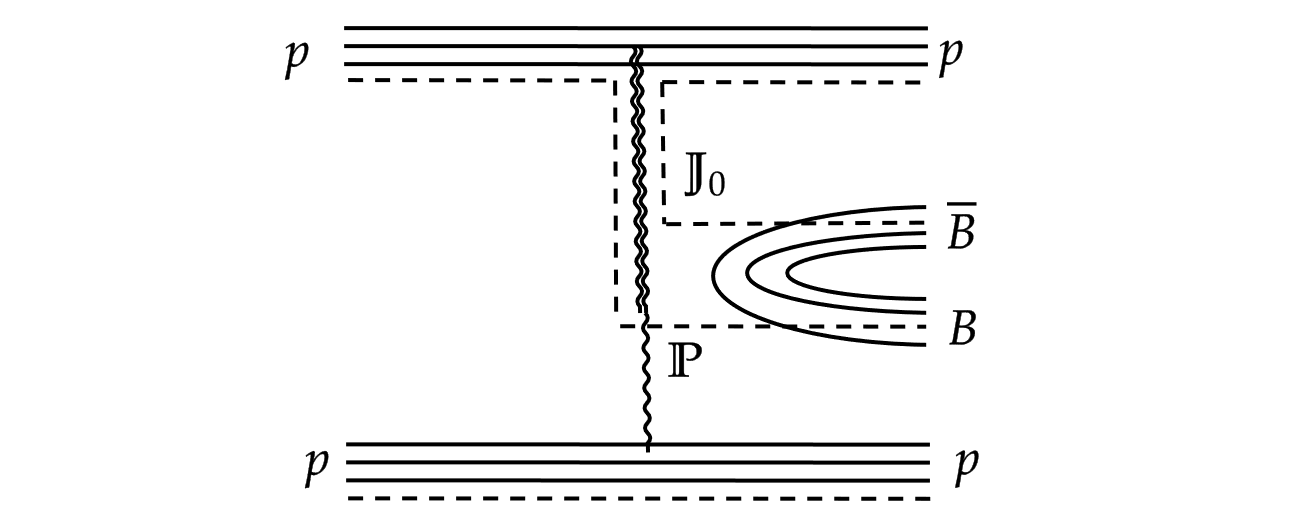}
    \caption{Double-diffractive production of $B\bar{B}$ pair that can be used to extract the slope of the $J_0$ trajectory.}
    \label{fig:BBbar_production}
\end{figure}

\item It will be useful to measure the rapidity distribution of $\bar B B$ pairs, as a function of rapidity separation between baryon and antibaryon. We expect that as a function of rapidity difference $\Delta y$ between the baryon and antibaryon, the distribution will behave as $\exp(-0.82\ \Delta y)$ at large $\Delta y$ due to the $J_0$ exchange. The corresponding associated hadron multiplicity (per unit rapidity) in the rapidity interval $\Delta y$ will be about $3/2$ of the average multiplicity in inclusive events. At smaller $\Delta y$, $J_2$ and $J_4$ trajectories will contribute, and provide terms proportional to $\exp(-1.32\ \Delta y)$ and $\exp(-1.74\ \Delta y)$, with associated multiplicity densities that are about $1$ and $1/2$ of the inclusive multiplicity, correspondingly.

\item It would be very important to measure the slope of the $J_0$ trajectory. One way of doing so is measuring the double-diffractive production of a baryon-antibaryon pair, as shown in Fig. \ref{fig:BBbar_production}. In a high-energy $pp$ collision, this would amount to detecting a baryon and an antibaryon at mid-rapidity that are separated by rapidity gaps from both of the final-state protons (or their excited states). In this process, the JOZI rule will suppress the contribution of two Pomerons that usually dominate the double-diffractive processes, and enhance the Pomeron - $J_0$ fusion amplitude, see Fig. \ref{fig:BBbar_production}. Measuring the double differential cross section of this process would thus allow to extract the slope of the $J_0$ trajectory.
\item Based on our prediction of degenerate $\cal C$-even and $\cal C$-odd $J_0$ trajectories we expect the existence of a heavy tensor $J^{{\cal PC}} = 2^{++}$ $J{\bar J}$ glueball with mass $M \simeq 3.0-4.1$ GeV, which is heavier than the $M \simeq 2.4$ GeV tensor glueball measured on the lattice \cite{Chen:2005mg}. We also predict $J\bar{J}$ glueballs with $J^{{\cal PC}} = 3^{-+}$ and $J^{{\cal PC}} = 3^{--}$ with masses in the range $3.8 - 5.2$ GeV. We suggest using correlations of junction-carrying lattice operators, such as the one in (\ref{eq:J0_lattice}), in order to enhance the sensitivity to these new states. 

\end{itemize}

In conclusion, the phenomenon of baryon-number--flavor separation points to the fundamental role played by the string junction in transporting the baryon number. It is important to explore this fascinating QCD phenomenon further, both experimentally and theoretically.  
\vspace{1cm}

{\large \bf Acknowledgments} This work was supported in part by the U.S. Department of Energy, Office of Science, Office of Nuclear Physics,   Grants No.
DE-FG88ER41450 (DF, DK) and DE-SC0012704
(DK). GCR acknowledges partial financial support from INFN IS Lcd123.

\appendix

\section{Derivation of equation (\ref{plSigma1})}
\setcounter{equation}{0}

In order to justify our starting point (\ref{plSigma1}) it is better to  
begin, as in \cite{Wilson:1970zzb}, from more differential (i.e. unintegrated) exclusive and  inclusive cross sections and their generating functionals. This will also be useful to describe rapidity distributions, as well as flavor and baryon-number transport over large rapidity intervals. We shall then recover (\ref{plSigma1}) upon integration over phase space.

We define the generating functional of exclusive cross sections by:
\beqn{Z}
\Sigma[z(x)] = \sum_n \int \prod_{j=1}^n (d x^j z(x^j)) \frac{1}{\sigma_t} \frac {d \sigma(a+b \ra x^1, x^2 \dots x^n )}{dx^1 d x^2 \dots dx^n}
\eeqn
where $x^j$ is a collective notation for  the ``coordinates"  of the $j^{th}$ final particle (i.e. whatever we want to measure about it: transverse momentum, rapidity, flavor, baryon number, etc.) and $d \sigma (x^1, x^2 \dots x^n)$ denotes the corresponding $n$-particle differential exclusive cross section, normalized to a suitably defined ``total" cross section $\sigma_t$ so that, by definition, $\Sigma[z(x)=1] =1$.

Clearly, individual exclusive differential cross section are given by partial functional differentiation of $\Sigma$ evaluated at $z(x) =0$. Instead, the $m$-particle inclusive cross section:
\beqn{rho}
\rho_m(x^1, x^2 \dots x^m) = \frac{1}{\sigma_t} \sum_X  \frac {d \sigma(a+b \ra x^1, x^2 \dots x^m  +X)}{dx^1 d x^2 \dots dx^m}
\eeqn
 is given by an order $m$ functional differentiation of  $\Sigma$ evaluated at $z(x) =1$. 
 
 A standard cluster-decomposition argument expresses those inclusive differential cross sections in terms of connected  correlators defined via the expansion of:
 \beqn{PV}
\log \Sigma[z(x)] =  \sum_m \frac{1}{m!} \int  \prod_{j=1}^m [d x^j (z(x^j)-1)] c_m (x^1, x^2 \dots x^m) \equiv p[z(x)] V
\eeqn
 around $z(x) =1$.
$\Sigma$, as defined in (\ref{Z}), is the analog of the grand-canonical partition function in statistical mechanics, whose logarithm is identified with $\frac{P V}{k_B T}$ after  having taken a large-$V$ thermodynamic limit.
 
 In our context we may consider $k_B T$ to be fixed in terms of the  average transverse momentum of the final particles while in (\ref{PV}) $V$ is a fixed ``volume factor" associated with the (log of the) total energy (see below) and $p[z(x)]$ plays the role of the pressure. Since we are interested in the high-energy limit, also in our case a large-$V$ limit is understood. Indeed it is important to define $p[z(x)]$ as the coefficient of $V$ in a large-$V$ expansion of $\log \Sigma[z(x)]$ at fixed $z$. This implies that $p[z(x)]$ is insensitive to a $V$-independent (but possibly $z$-dependent) rescaling of $\Sigma$. For instance,  one may like to add  a $z^{-2}$ overall factor  in order to make $\Sigma$ approach a finite limit for  $z \to 0$, without affecting $p[z(x)]$.

 The first few terms of the expansion give:
\beqn{CD}
\hspace{-1.4cm}&&\rho_1(x) = c_1(x)  \nonumber \\
\hspace{-1.4cm}&&\rho_2(x^1,x^2)= c_1(x^1) c_1(x^2) + c_2(x^1,x^2)   \nonumber \\
\hspace{-1.4cm}&&\rho_3(x^1,x^2,x^3)= c_1(x^1) c_1(x^2)c_1(x^3) + c_2(x^1,x^2) c_1(x^3) + {\mbox{perm.}}
+ c_3(x^1,x^2,x^3)  \nonumber \\ 
\hspace{-1.4cm}&&  \rho_4(x^1,x^2,x^3,x^4)= c_1(x^1) c_1(x^2)c_1(x^3)c_1(x^4) + c_1(x^1) c_1(x^2)c_2(x^3,x^4) + {\mbox{perm.}}\nonumber \\ 
 \hspace{-1.4cm}&& \quad + c_3(x^1,x^2,x^3) c_1(x^4)+ {\mbox{perm.}} + c_2(x^1,x^2) c_2(x^3, x^4)+ {\mbox{perm.}}
 + c_4(x^1,x^2,x^3,x^4)\dots
\eeqn
 Let us now comment on the relation between (\ref{CD}) and some equations in the main text. The first point is that, on the basis of a Mueller-Kancheli analysis \cite{Mueller:1970fa,Kancheli:1970gt} one can argue that the correlators $c_m (x^1, x^2 \dots x^m) $ are short range in the sense that that they fall of rapidly whenever any two subsets of the $x^i$ get separated by a large distance in rapidity. Assuming also, that there is a finite transverse momentum cut off, integrating any $c_m$ over it arguments will produce just single power of the total rapidity interval $Y$. This is to be contrasted with the integral of total correlators $\rho_m$ which will typically grow as $Y^m$. So a term in the expansion (\ref{CD}) of $\rho_m$ with a number $q$ of connected correlators $c$ will give a contribution of order $Y^q$ (this is how the various contributions are ordered in (\ref{CD}) ).
 The second observation is the well known fact (see e.g.\ \cite{DeTar:1971pmj}) that the total integral of $\rho_m$ gives the expectation value  $\langle n(n-1)\dots(n-m+1)\rangle$ explaining why we get such factors in the integrated form  (\ref{plSigma1}) of (\ref{CD}). We remark
 that the numerical coefficients appearing in (\ref{plSigma1}) correspond exactly to the number of distinct permutations  
 of each structure appearing in (\ref{CD}).
 
 The extension of this procedure to the case of several species is rather straightforward and leads to results similar to those of (\ref{CD}). The main difference is that now we want to distinguish correlation within one species from correlation among 
 different species. Then each term in the cluster decomposition will 
 appear with the appropriate weight coming from the number of distinct  ways to group the particles in separate clusters within and across different species.

\bibliographystyle{ytphys}
\bibliography{main}

\end{document}